\documentclass[draftclsnofoot, onecolumn, 12pt]{IEEEtran}
\usepackage{xcolor}
\usepackage{amsmath,graphicx}
\usepackage{amssymb}
\usepackage{subfigure}
\usepackage{footmisc}
\usepackage{lipsum}
\usepackage{mathtools}
\usepackage{cuted}
\usepackage{esint}
\usepackage{cite}
\usepackage{etoolbox}

\makeatletter
\@for\next:={int,iint,iiint,iiiint,dotsint,oint,oiint,sqint,sqiint,
  ointctrclockwise,ointclockwise,varointclockwise,varointctrclockwise,
  fint,varoiint,landupint,landdownint}\do{%
    \expandafter\edef\csname\next\endcsname{%
      \noexpand\DOTSI
      \expandafter\noexpand\csname\next op\endcsname
      \noexpand\ilimits@
    }%
  }
\makeatother

\begin{document}

\title{Performance Analysis of Large Intelligent Surfaces (LISs): Asymptotic Data Rate and Channel Hardening Effects}

\author{Minchae~Jung,~\IEEEmembership{Member,~IEEE}, Walid~Saad,~\IEEEmembership{Fellow,~IEEE}, \\
Youngrok~Jang, Gyuyeol~Kong,~\IEEEmembership{Member,~IEEE},
and Sooyong~Choi,~\IEEEmembership{Member,~IEEE}

\thanks{
A preliminary version of this work was submitted to EuCNC 2020 \cite{ref.EuCNC}.

{M. Jung and W. Saad are with Wireless@VT, Department of Electrical and Computer Engineering, Virginia Tech, Blacksburg, VA 24061 USA (e-mail: hosaly@vt.edu; walids@vt.edu).}

{Y. Jang, G. Kong, and S. Choi are with School of Electrical Electronic Engineering, Yonsei University, Seoul 03722, Korea (e-mail: dynamics@yonsei.ac.kr; gykong@yonsei.ac.kr; csyong@yonsei.ac.kr).}

This research was supported by Basic Science Research Program through the National Research Foundation of Korea (NRF) funded by the Ministry of Education (NRF-2016R1A6A3A11936259)
and by the U.S. National Science Foundation under Grants IIS-1633363 and OAC-1638283.
\vspace{-0.3cm}
}
}

\markboth{}%
{Shell \MakeLowercase{\textit{et al.}}: Bare Demo of IEEEtran.cls for Journals}

\maketitle

\vspace{-1.8cm}
\begin{abstract}
\vspace{-0.3cm}
The concept of a large intelligent surface (LIS) has recently emerged as a promising wireless communication paradigm that can exploit the entire surface of man-made structures for transmitting and receiving information.
An LIS is expected to go beyond massive multiple-input multiple-output (MIMO) system, insofar as the desired channel can be modeled as a perfect line-of-sight.
To understand the fundamental performance benefits, it is imperative to analyze its achievable data rate, under practical LIS environments and limitations.
In this paper, an asymptotic analysis of the uplink data rate in an LIS-based large antenna-array system is presented. 
In particular, the asymptotic LIS rate is derived in a practical wireless environment where the estimated channel on LIS is subject to estimation errors, interference channels are spatially correlated Rician fading channels, {and the LIS experiences hardware impairments.}
Moreover, the occurrence of the channel hardening effect is analyzed and the performance bound is asymptotically derived for the considered LIS system. 
The analytical asymptotic results are then shown to be in close agreement with the exact mutual information as the number of antennas and devices increase without bounds. 
Moreover, the derived ergodic rates show that {hardware impairments,} noise, and interference from estimation errors and the non-line-of-sight path become negligible as the number of antennas increases. 
Simulation results show that an LIS can achieve a performance that is comparable to conventional massive MIMO with improved reliability and a significantly reduced area for antenna deployment.
\end{abstract}

\vspace{-0.5cm}
\begin{IEEEkeywords}
\vspace{-0.3cm}
large intelligent surface (LIS), large system analysis, channel estimation, ergodic rate, channel hardening effect.

\end{IEEEkeywords}

\IEEEpeerreviewmaketitle
\vspace{-0.5cm}
\section{Introduction}
\IEEEPARstart{F}{uture} man-made structures, such as buildings, roads, and walls, are expected to be electromagnetically active \cite{ref.LIS1,ref.LIS2,ref.LIS3,ref.LIS4}.
As such, these structures can be leveraged to provide wireless connectivity to emerging services, such as Internet of Things (IoT) applications \cite{ref.Walid20196G,ref.IoT,ref.Saad1,ref.Saad2,ref.Saad3,ref.Saad4}, via the emerging concept of a large intelligent surface (LIS) \cite{ref.LIS1,ref.LIS2,ref.LIS3}.
If properly operated and deployed, LISs are expected to provide wireless connectivity to a plethora of IoT devices, such as sensors, vehicles, and surveillance cameras, through man-made structures.
The LIS concept can be essentially viewed as a scaled-up version of conventional massive multiple-input and multiple-output (MIMO) systems.
However, an LIS exhibits several key differences from massive MIMO systems.
First, unlike conventional massive MIMO systems where transmission and reception are carried out via a base station (BS), 
an LIS can transmit and receive signals through all surfaces of man-made structures. 
{
This allows users in close proximity to communicate with an LIS and their transmission power levels can be set to values that are lower than those resulting from massive MIMO.
This results in battery savings at the device and reduced interference levels in an LIS.
Hence, higher data rates can be achieved because of the reduced interference levels, compared to massive MIMO systems.}
%
Second, LISs will be densely located in both indoor and outdoor spaces, making it possible to perform near-field communications through a line-of-sight (LOS) path \cite{ref.LIS1,ref.LIS2,ref.LIS3}.
Since an LOS path is highly correlated to the channel components between antennas, antenna spacing of greater than half a wavelength is meaningless in order to obtain full diversity gain.
Consequently, an LIS can enable dense antenna arrays\footnote{Indeed, distortions in radiation patterns can occur due to mutual coupling. 
However, we envision an LIS that can correct the distortions in the radiation patterns of any array with any antenna spacing, as in \cite{ref.Coupling1} and \cite{ref.Coupling2}.
More practically, we consider hardware impairments which include residual coupling loss after decoupling of antenna-RF chains as in \cite{ref.Christoph2010HWI}.}.
Finally, an LIS enables simpler channel estimation and feedback, compared to conventional massive MIMO systems
that typically require channel state information (CSI) for hundreds of antennas.
Since an extensive overhead for CSI acquisition resulting from pilot training and CSI feedback can be caused by the massive number of antennas, this overhead can seriously degrade the performance of massive MIMO systems \cite{ref.FB1}, \cite{ref.FB2}.
However, the desired channel of an LIS-based large antenna-array system is highly correlated with the LOS path, 
facilitating accuracy and simplicity in terms of channel estimation and feedback.

For these reasons, the use of an LIS has recently attracted attention in the wireless literature \cite{ref.LIS1,ref.LIS2,ref.LIS3,ref.LIS4}.
These recent works focus on addressing a number of LIS challenges that include performance analysis, estimating user location, user assignment, and power allocation. 
For instance, in \cite{ref.LIS1}, the authors derived the data rates of the optimal receiver and the matched filter (MF) in the uplink of an LIS-based system.
Meanwhile, in \cite{ref.LIS2}, the authors obtained the Fisher-information and Cramer--Rao lower bound for user positions using the uplink signal for an LIS. 
In \cite{ref.LIS3}, the authors proposed optimal user assignments to select LIS units that maximize the sum rate and minimum individual rate.
The authors in \cite{ref.LIS4} proposed the use of LIS as a relay station for a massive MIMO system and developed a power allocation scheme to maximize energy efficiency.
However, these previous studies have not considered practical LIS environments and their limitations, such as imperfect channel estimations and a user-specific channel model.
For instance, both \cite{ref.LIS1} and \cite{ref.LIS2} assumed an LIS with an infinite surface area and considered that a single infinite LIS performs the MF over all devices. 
Moreover, \cite{ref.LIS1,ref.LIS2,ref.LIS3,ref.LIS4} assumed perfect channel estimations for an LIS. 
Finally, all of the interference channels in \cite{ref.LIS1,ref.LIS2,ref.LIS3} were assumed as following a LOS path, and \cite{ref.LIS4} considered independent Rayleigh fading both in desired and interference channels. 
Given that LISs are densely located and devices are reasonably close to their target LISs, desired channels can be modeled as a LOS path whereas interference channels must be modeled depending on the distances between interfering devices and the target LISs.
Therefore, the interference channel can be composed of a deterministic LOS path and spatially correlated non-line-of-sight (NLOS) path describing a device-specific spatially correlated multipath environment.

The main contribution of this paper is a rigorous asymptotic analysis of the uplink rate of an LIS-based large antenna-array system that considers a practical LIS environment and its limitations. 
In this regard, we assume that each device uses as desired surface that maps to a limited area of the entire LIS that we refer to as an LIS unit.
Further, the MF procedure across the surface is assumed to be performed under \emph{realistic channel estimation errors}
and \emph{hardware impairments}, such as analog imperfectness, quantization errors, and residual coupling loss, are also considered. 
The interference channels are modeled as spatially correlated Rician fading channels, composed of a deterministic LOS path and stochastic NLOS path according to the distance between the interfering device and the target LIS unit. 
We then analyze the uplink ergodic rate of each device in presence of a large number of antennas and devices, and derive an asymptotic ergodic rate of LIS.
This approximation allows the estimation of the uplink ergodic rate accurately without the need for extensive simulations,
and then it enables to obtain optimal operating parameters such as an optimal size of an LIS unit.
The asymptotic variance of the uplink rate is also derived in order to verify the occurrence of \emph{channel hardening effect} theoretically,
that is a particularly important phenomenon in large antenna-array systems such as massive MIMO and LIS \cite{ref.Hard}.
Given that the channel hardening determines several practical implications such as system reliability, latency, and scheduling diversity, 
we analyze the occurrence of the channel hardening effect in an LIS-based system and compare it to a massive MIMO system.
On the basis of the asymptotic ergodic rate and variance, the performance bound of an LIS-based system is obtained by using a scaling law for the uplink signal-to-interference-plus-noise ratio (SINR).
We then show a particular operating characteristic of LIS whereby noise, estimation errors, hardware impairments, and NLOS interference become negligible compared to LOS interference from other devices.
Our simulations show that LIS can be a promising technology beyond massive MIMO given that LIS can provide a comparable rate to massive MIMO, with improved reliability and a significantly reduced area for antenna deployment.

The rest of this paper is organized as follows. 
Section II presents the LIS-based system model. 
Section III describes the asymptotic analysis of the uplink data rate
and Section IV describes the channel hardening effect and performance bound. 
Simulation results are provided in Section V to support and verify the analyses, and
Section VI concludes the paper.

\textit{Notations:} Throughout this paper, boldface upper- and lower-case symbols represent matrices and vectors respectively, and $\boldsymbol{I}_M$ denotes a size-$M$ identity matrix.
The conjugate, transpose, and Hermitian transpose operators are denoted by  ${\left(  \cdot  \right)^*}$, ${\left(  \cdot  \right)^{\rm{T}}}$, and ${\left(  \cdot  \right)^{\rm{H}}}$, respectively. 
The norm of a vector $\boldsymbol{a}$ is denoted by  $\left| {\boldsymbol{a}} \right|$.
\begin{figure}[!ht]
\centering
\includegraphics[width=0.69\columnwidth] {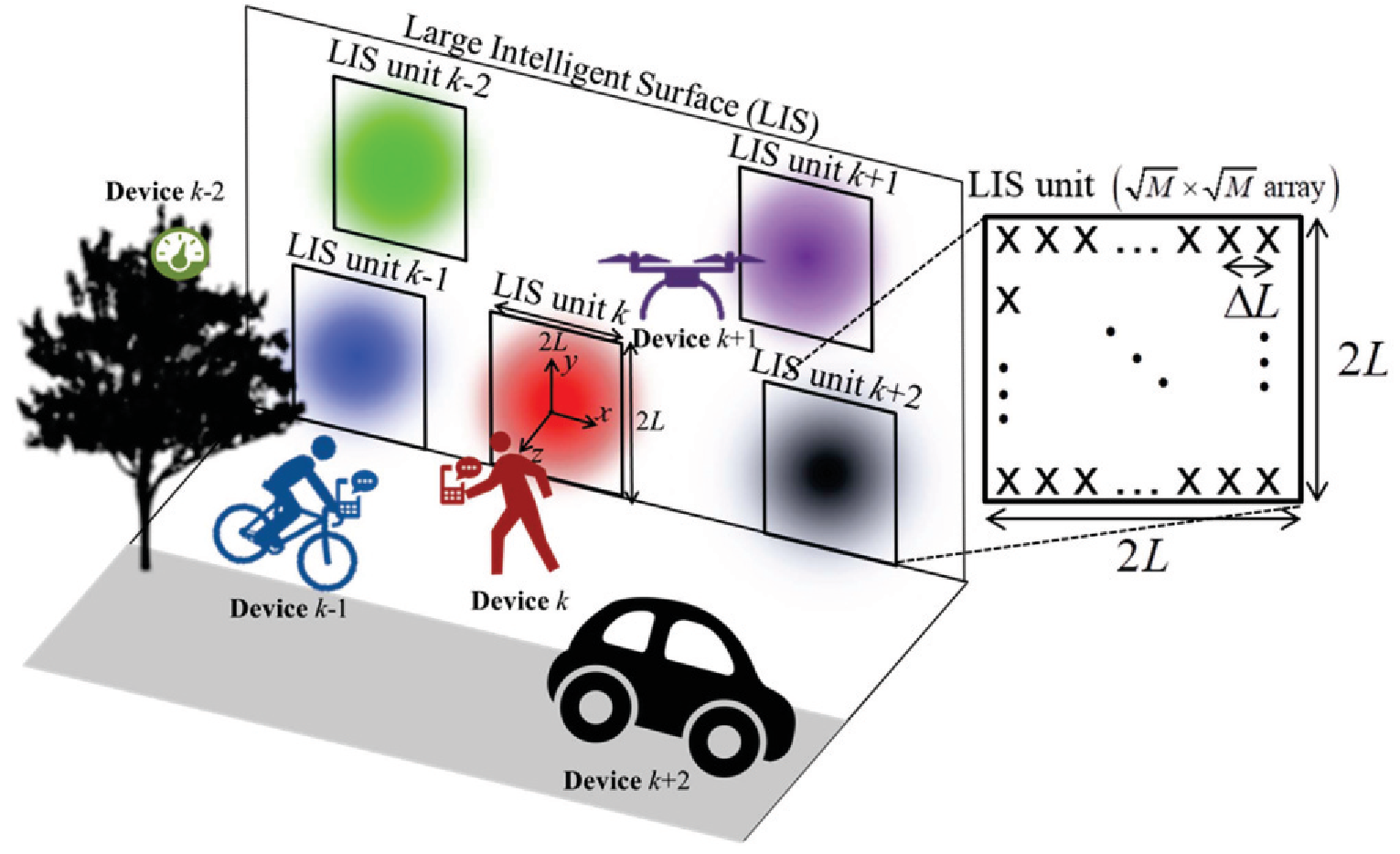}
\caption{Illustrative system model of the considered uplink LIS having multiple LIS units and serving $K$ users.}
\label{fig.1}
\end{figure}
${\rm{E}}\left[  \cdot  \right]$, ${\rm{Var}}\left[  \cdot  \right]$, and ${\rm{Cov}}\left[  \cdot  \right]$ denote expectation, variance, and covariance operators, respectively.
${\mu _X}={\rm{E}}\left[  X  \right]$ and $\sigma _X^2={\rm{Var}}\left[  X  \right]$ denote the mean and variance of a random variable $X$, respectively.
$\mathcal{O}\left(  \cdot  \right)$, $ \otimes$, and $\circ$ denote the big O notation, the Kronecker product, and the Hadamard product, respectively.
The operators ${\mathop{\rm Re}\nolimits} \left(  \cdot  \right)$ take the real part.
 $\mathcal{CN}\left( {m,{\sigma ^2}} \right)$ and $\chi _k^2$ denote a complex Gaussian distribution with mean $m$ and variance $\sigma ^2$, and a chi-square distribution with $k$ degrees of freedom, respectively.

\section{System Model}
We consider an uplink LIS-based large antenna-array system that serves $K$ single-antenna devices, as shown in Fig. \ref{fig.1}. 
The LIS is located in a two-dimensional space along the $xy$-plane at $z = 0$ in Cartesian coordinates. 
We define the notion of an \emph{LIS unit} which corresponds to a subarea of the entire LIS and
has a square shape with an area limited to $2L \times 2L$ centered on the $(x, y)$ coordinates of the corresponding device. 
Each LIS unit has a large number of antennas, $M$, distributed on its surface with antenna spacing of $\Delta L$ in a rectangular lattice form. 
We assume that each LIS unit has its own signal processing unit for estimating the channel and detecting
any data signal, as in \cite{ref.LIS1,ref.LIS2,ref.LIS3}, \cite{ref.Black2017holographic}, and \cite{ref.Tang2018wireless}.
Each user device communicates with its corresponding LIS unit and controls the transmission power toward the center of the LIS unit according to target signal-to-noise-ratio (SNR), in order to avoid the near-far problem. 
We define the location of device $k$ as $({x_k},{y_k},{z_k})$.
Then, antenna $m$ of LIS unit $k$ will be located at $({x^{\rm{LIS}}_{km}},{y^{\rm{LIS}}_{km}},0)$ where $x_{km}^{{\rm{LIS}}} \in \left[ {{x_k} - L,{x_k} + L} \right]$ and $y_{km}^{{\rm{LIS}}} \in \left[ {{y_k} - L,{y_k} + L} \right]$.
In contrast to the works in \cite{ref.LIS1} and \cite{ref.LIS2}, that consider an infinite $L$, we consider a finite $L$ which is a more practical assumption.


Depending on the location of the device, LIS units may overlap which, in turn, can seriously degrade the performance. 
To overcome this problem, we assume that the LIS allocates orthogonal resources among devices with similar $(x, y)$ coordinates using an appropriate resource allocation and scheduling scheme. 
Therefore, we assume that each device communicates with a non-overlapping LIS unit.

\subsection{Wireless Channel Model}
The desired channel ${{\boldsymbol{h}}_{kk}} \in \mathbb{C}{^M}$ between device $k$ and LIS unit $k$ is assumed to be a LOS path. 
Then, ${{\boldsymbol{h}}_{kk}} $ can be given by 
${{\boldsymbol{h}}_{kk}}  = {[ {\beta _{kk1}^{\rm{L}}{h_{kk1}}, \ldots ,\beta _{kkM}^{\rm{L}}{h_{kkM}}} ]^{\rm{T}}},$ 
where $\beta _{kkm}^{\rm{L}} = \alpha _{kkm}^{\rm{L}}l_{kkm}^{\rm{L}}$ denotes a LOS channel gain between device $k$ and antenna $m$ of LIS unit $k$.
Here, $\alpha_{kkm}^{\rm{L}} = \sqrt {\cos {\theta _{kkm}}} $ denotes the antenna gain and  ${\theta _{kkm}}$ is the {azimuth angle-of-arrival} between device $k$ and antenna $m$ of LIS unit $k$. 
Given that the antennas on each LIS unit are placed at different locations within the $2L \times 2L$ square-shaped area, 
${\theta _{kkm}}$ has different values for different $m$ considering the non-isotropic characteristic of an LIS.
Since device $k$ has a distance of ${z_k}$  from the center of its target LIS unit, we obtain $\cos {\theta _{kkm}} = z_k/d_{kkm}$, where $d_{kkm}$ denotes the distance between device $k$ and antenna $m$ of LIS unit $k$, given by
${d_{kkm}} = \sqrt {{{( {{x_k} - {{x}^{\rm{LIS}}_{km}}} )}^2} + {{( {{y_k} - {{y}^{\rm{LIS}}_{km}}} )}^2} + z_k^2}.$
Also, $l_{kkm}^{\rm{L}} = 1/\sqrt {4\pi d_{kkm}^2}$ is the free space path loss attenuation,
and $h_{kkm}$ is the LOS channel state between device $k$ and antenna $m$ of LIS unit $k$, obtained as
${h_{kkm}} = \exp \left( { - j2\pi {d_{kkm}}/\lambda } \right)$, 
where $\lambda$ denotes the wavelength of a signal \cite{ref.TSE}.
In fact, the desired channel can be generated by Rician fading composed of a deterministic LOS path and spatially correlated NLOS path.
However, the signal from the NLOS path becomes negligible compared to the one from the LOS path as $M$ increases, as will be proved in Section IV.
Therefore, the desired channels are modeled as a \emph{perfect LOS path} in the considered LIS as assumed in many prior works \cite{ref.LIS1,ref.LIS2,ref.LIS3}.
Further, ${{\boldsymbol{h}}_{jk}} \in \mathbb{C}{^M}$ is the interference channel between device $j$ and LIS unit $k$, expressed as a combination of LOS and NLOS:
\begin{equation}
{{\boldsymbol{h}}_{jk}} = \sqrt {\frac{{{\kappa _{jk}}}}{{{\kappa _{jk}} + 1}}} {\boldsymbol{h}}_{jk}^{\rm{L}} + \sqrt {\frac{1}{{{\kappa _{jk}} + 1}}} {\boldsymbol{h}}_{jk}^{{\rm{NL}}}, 
\label{eq.hjk}
\end{equation}
where $\kappa _{jk}$ is the Rician factor between device $j$ and LIS unit $k$.  
Here, ${\boldsymbol{h}}_{jk}^{\rm{L}} \in \mathbb{C}{^M}$ is the deterministic LOS component from device $j$ to LIS unit $k$ given by
${\boldsymbol{h}}_{jk}^{\rm{L}} ={\left[ {\beta _{j1}^{\rm{L}}{h_{jk1}}, \ldots ,\beta _{jM}^{\rm{L}}{h_{jkM}}} \right]^{\rm{T}}}$,
where $\beta _{jm}^{\rm{L}} = \alpha _{jkm}^{\rm{L}}l_{jkm}^{\rm{L}}$ and $h_{jkm}$ are LOS channel gain and state, respectively, between device $j$ and antenna $m$ of LIS unit $k$.
${\boldsymbol{h}}_{jk}^{\rm{NL}}$ is the correlated NLOS component defined as
${\boldsymbol{h}}_{jk}^{{\rm{NL}}} = {\left[ {{h_{jk1}^{\rm{NL}}}, \ldots ,{h_{jkM}^{\rm{NL}}}} \right]^{\rm{T}}}
={\boldsymbol{R}}_{jk}^{1/2}{{\boldsymbol{g}}_{jk}}$, 
where ${{\boldsymbol{R}}_{jk}} \in \mathbb{C}{^{M \times P}}$ is the deterministic correlation matrix from device $j$ to LIS unit $k$
and ${{\boldsymbol{g}}_{jk}} = {\left[ {{g_{jk1}}, \ldots ,{g_{jkP}}} \right]^{\rm{T}}} \sim \mathcal{CN}\left( {{\boldsymbol{0}},{{\boldsymbol{I}}_P}} \right)$ is an independent fast-fading channel vector.
Here, $P$ represents the number of dominant paths among all NLOS paths and is related to the amount of scattering in the wireless channel environment \cite{ref.How}. 
Since we consider an LIS located in a two-dimensional space along the $xy$-plane, it can be modeled as a uniform planar array (UPA) \cite{ref.UPA}.
Given a UPA model, ${{\boldsymbol{R}}_{jk}}$ is obtained as
${\boldsymbol{R}}_{jk}^{1/2} = {\boldsymbol{l}}_{jk}^{{\rm{NL}}}{{\boldsymbol{D}}_{jk}}$, where
${{\boldsymbol{D}}_{jk}} = [ {\alpha _{jk1}^{{\rm{NL}}}{\boldsymbol{d}}( {\phi _{jk1}^{\rm{v}},\phi _{jk1}^{\rm{h}}} ), \ldots ,\alpha _{jkP}^{{\rm{NL}}}{\boldsymbol{d}}( {\phi _{jkP}^{\rm{v}},\phi _{jkP}^{\rm{h}}} )} ]$
and ${\boldsymbol{l}}_{jk}^{{\rm{NL}}} = {\rm{diag}}( {l_{jk1}^{{\rm{NL}}}  \ldots ,l_{jkM}^{{\rm{NL}}}} )$ is a diagonal matrix including the path loss attenuation factors $l_{jkm}^{{\rm{NL}}} = d_{jkm}^{ - {\beta _{{\rm{PL}}}}/2}$ with a path loss exponent ${\beta _{{\rm{PL}}}}$.
Here, ${\boldsymbol{d}}\left( {\phi _{jkp}^{\rm{v}},\phi _{jkp}^{\rm{h}}} \right) \in \mathbb{C}{^M}$ represents NLOS path $p$ at given angles of $\left( {\phi _{jkp}^{\rm{v}},\phi _{jkp}^{\rm{h}}} \right)$.
By using a UPA model, ${\boldsymbol{d}}\left( {\phi _{jkp}^{\rm{v}},\phi _{jkp}^{\rm{h}}} \right)$ will be given by:
\begin{gather}
{\boldsymbol{d}}\left( {\phi _{jkp}^{\rm{v}},\phi _{jkp}^{\rm{h}}} \right) = \frac{1}{{\sqrt M }}{{\boldsymbol{d}}_{\rm{v}}}\left( {\phi _{jkp}^{\rm{v}}} \right) \otimes {{\boldsymbol{d}}_{\rm{h}}}\left( {\phi _{jkp}^{\rm{h}}} \right), \label{eq.DNLOS}\\
{{\boldsymbol{d}}_{\rm{v}}}\left( {\phi _{jkp}^{\rm{v}}} \right) = {\left[ {1,{\rm{ }}{e^{j\frac{{2\pi \Delta L}}{\lambda }\phi _{jkp}^{\rm{v}}}}, \cdots ,{\rm{ }}{e^{j\frac{{2\pi \Delta L}}{\lambda }\left( {\sqrt M  - 1} \right)\phi _{jkp}^{\rm{v}}}}} \right]^{\rm{T}}}, \\
{{\boldsymbol{d}}_{\rm{h}}}\left( {\phi _{jkp}^{\rm{h}}} \right) = {\left[ {1,{\rm{ }}{e^{j\frac{{2\pi \Delta L}}{\lambda }\phi _{jkp}^{\rm{h}}}}, \cdots ,{\rm{ }}{e^{j\frac{{2\pi \Delta L}}{\lambda }\left( {\sqrt M  - 1} \right)\phi _{jkp}^{\rm{h}}}}} \right]^{\rm{T}}}, 
\end{gather}
where $\phi _{jkp}^{\rm{v}} = \sin \theta _{jkp}^{\rm{v}}$ and $\phi _{jkp}^{\rm{h}} = \sin \theta _{jkp}^{\rm{h}}\cos \theta _{jkp}^{\rm{v}}$ when the elevation and azimuth angles of path $p$ between device $j$ and LIS unit $k$
are $\theta _{jkp}^{\rm{v}}$ and $\theta _{jkp}^{\rm{h}}$, respectively \cite{ref.UPAbasic}. 
$\alpha _{jkp}^{{\rm{NL}}}$ indicates the antenna gain of path $p$ which can be obtained by $\alpha _{jkp}^{{\rm{NL}}} = \sqrt {\cos \theta _{jkp}^{\rm{v}}\cos \theta _{jkp}^{\rm{h}}}$
where ${\theta _{jkp}} \in \left[ { - \frac{\pi }{2},\frac{\pi }{2}} \right]$ and ${\theta _{jkp}} \in \left\{ {\theta _{jkp}^{\rm{v}},\theta _{jkp}^{\rm{h}}} \right\}$.

\subsection{Uplink Data Rate}
The instantaneous uplink data rate of device $k$ will be given by
${R_k} = \log \left( {1 + {\gamma _k}} \right),$ 
where ${\gamma _k}$ is the instantaneous SINR of device $k$ received at LIS unit $k$.
The uplink signal received from all devices at LIS unit $k$ is expressed as follows:
\begin{equation}
{{\boldsymbol{y}}_k} = \sqrt {{\rho_k}} {{\boldsymbol{h}}_{kk}}{x_k} + \sum\limits_{j \ne k}^K {\sqrt {{\rho_j}} {{\boldsymbol{h}}_{jk}}{x_j}}  + {{\boldsymbol{w}}_k} + {{\boldsymbol{n}}_k},
\end{equation} where ${x_k}$ and ${x_j}$ are uplink transmit signals of devices $k$ and $j$, respectively, assumed as independent Gaussian variables with zero means and unit variances.
Further, ${\rho_k}$ and ${\rho_j}$ are the uplink transmit SNRs of devices $k$ and $j$, respectively, and ${{\boldsymbol{n}}_k} \in\mathbb{C} {^M} \sim \mathcal{CN}\left( {{\boldsymbol{0}},{{\boldsymbol{I}}_M}} \right)$ is the noise vector. 
Moreover, ${{\boldsymbol{w}}_k}\in\mathbb{C} {^M}$ represents the residual noise caused by hardware impairments, as given by
\begin{equation}
{{\boldsymbol{w}}_k} = {{\boldsymbol{c}}_k} \circ \left( {\sqrt {{\rho _k}} {{\boldsymbol{h}}_{kk}}{x_k} + \sum\limits_{j \ne k}^K {\sqrt {{\rho _j}} {{\boldsymbol{h}}_{jk}}{x_j}} } \right),\label{eq.HWI}
\end{equation}
where ${{\boldsymbol{c}}_k}={\left[ c_1, \ldots ,c_M \right]^{\rm{T}}}$ represents hardware impairments at LIS unit $k$ which can be modeled using a Gaussian distribution such that ${{\boldsymbol{c}}_k}\sim \mathcal{CN}\left( {{\boldsymbol{0}},\delta{{\boldsymbol{I}}_M}} \right)$ \cite{ref.Sha2018HWI,ref.Christoph2010HWI}.
Therefore, in (5),  $\sqrt {{\rho_k}} {{\boldsymbol{h}}_{kk}}{x_k}$ is  uplink desired signal of device $k$, 
$\sum\nolimits_{j \ne k}^K {\sqrt {{\rho_j}} {{\boldsymbol{h}}_{jk}}{x_j}}$ is the aggregate uplink interference from other devices,
and ${{\boldsymbol{w}}_k} + {{\boldsymbol{n}}_k}$ is the sum of the noise components.
We consider a linear receiver ${\boldsymbol{f}}_k^{\rm{H}}$ for signal detection.
Then, the received signal at LIS unit $k$ is obtained as
\begin{equation}
{\boldsymbol{f}}_k^{\rm{H}}{{\boldsymbol{y}}_k} = \sqrt {{\rho_k}} {\boldsymbol{f}}_k^{\rm{H}}{{\boldsymbol{h}}_{kk}}{x_k} + \sum\limits_{j \ne k}^K {\sqrt {{\rho_j}} {\boldsymbol{f}}_k^{\rm{H}}{{\boldsymbol{h}}_{jk}}{x_j}}  + {\boldsymbol{f}}_k^{\rm{H}}{{\boldsymbol{w}}_k}+ {\boldsymbol{f}}_k^{\rm{H}}{{\boldsymbol{n}}_k}. \label{eq.fyk}
\end{equation}
We consider an MF receiver defined by ${{\boldsymbol{f}}_k} = {{\boldsymbol{\hat h}}_{kk}}$ where ${{\boldsymbol{\hat h}}_{kk}}$ is the estimated channel of ${{\boldsymbol{h}}_{kk}}$. 
In the case of perfect channel estimation, ${{\boldsymbol{f}}_k} = {{\boldsymbol{h}}_{kk}}$. 
Under the imperfect CSI results from an least square estimator, we have 
${{\boldsymbol{f}}_k} = {{\boldsymbol{h}}_{kk}} + \sqrt{\frac{\tau^2}{{1-\tau^2}}}{{\boldsymbol{e}}_k},$ 
where parameter ${\tau _k} \in [0,1]$ represents the imperfectness of ${{\boldsymbol{\hat h}}_{kk}}$.
${{\boldsymbol{e}}_k} = {\left[ {\beta _{k1}^{\rm{L}}{e_{k1}},...,\beta _{kM}^{\rm{L}}{e_{kM}}} \right]^{\rm{T}}} \in \mathbb{C}{^M}$  denotes the estimation error vector uncorrelated with ${{\boldsymbol{h}}_{kk}}$ and ${{\boldsymbol{n}}}_{k}$.
The elements of the estimation error vector has independent random variables of ${e_{km}} \sim \mathcal{CN} \left( {0,1} \right)$.
Using (\ref{eq.fyk}), we can write the received SINR of device $k$ at LIS unit $k$ as follows:
\begin{equation}
{\gamma _k} = \frac{{{\rho _k}\left( {1 - \tau _k^2} \right){{\left| {{{\boldsymbol{h}}_{kk}}} \right|}^4}}}{{{\rho _k}\tau _k^2{{\left| {{\boldsymbol{e}}_k^{\rm{H}}{{\boldsymbol{h}}_{kk}}} \right|}^2} + \sum\limits_{j \ne k}^K {{\rho _j}{{\left| {\sqrt {1 - \tau _k^2} {\boldsymbol{h}}_{kk}^{\rm{H}}{{\boldsymbol{h}}_{jk}} + {\tau _k}{\boldsymbol{e}}_k^{\rm{H}}{{\boldsymbol{h}}_{jk}}} \right|}^2}}  +\left|{\boldsymbol{f}}_k^{\rm{H}}{{\boldsymbol{\widetilde w}}_k}\right|^2
+{{\left| {\sqrt {1 - \tau _k^2} {\boldsymbol{h}}_{kk}^{\rm{H}} + {\tau _k}{\boldsymbol{e}}_k^{\rm{H}}} \right|}^2}}}, \label{eq.SINR}
\end{equation}
where 
${{\boldsymbol{\widetilde w}}_k}={\left[ {\widetilde w}_1, \ldots ,{\widetilde w}_M \right]^{\rm{T}}} = {{\boldsymbol{c}}_k} \circ  {\sum\nolimits_{i=1}^K {\sqrt {{\rho _i}} {{\boldsymbol{h}}_{ik}}} }$ .
Then, (\ref{eq.SINR}) can be simplified to
\begin{equation}
{\gamma _k} 
= \frac{{{\rho _k}{S_k}\left( {1 - \tau _k^2} \right)}}{{{I_k}}}, \label{eq.simpleSINR2}
\end{equation}
where
\begin{align}
{S_k} &= {{\left| {{{\boldsymbol{h}}_{kk}}} \right|}^4}, \forall k\label{eq.Sk_def}\\
{I_k} &= {{{\rho _k}\tau _k^2{X_k} + \sum\limits_{j \ne k}^K {{\rho _j}{Y_{jk}}}  + {Z_k}}}, \forall k \label{eq.Ik_def}
\end{align}
and
\begin{align}
{X_k} &= {{\left| {{\boldsymbol{e}}_k^{\rm{H}}{{\boldsymbol{h}}_{kk}}} \right|}^2}, \forall k \label{eq.Xk}\\
{Y_{jk}} &= {{\left| {\sqrt {1 - \tau _k^2} {\boldsymbol{h}}_{kk}^{\rm{H}}{{\boldsymbol{h}}_{jk}} + {\tau _k}{\boldsymbol{e}}_k^{\rm{H}}{{\boldsymbol{h}}_{jk}}} \right|}^2}, \forall j,k \label{eq.Yjk}\\
{{Z_k}} & {= Z_k^{\rm{w}}+Z_k^{\rm{n}}
={{\left| {\sqrt {1 - \tau _k^2} {\boldsymbol{h}}_{kk}^{\rm{H}}{{\boldsymbol{\widetilde w}}_k} + {\tau _k}{\boldsymbol{e}}_k^{\rm{H}}}{{\boldsymbol{\widetilde w}}_k} \right|}^2}+{{\left| {\sqrt {1 - \tau _k^2} {\boldsymbol{h}}_{kk}^{\rm{H}} + {\tau _k}{\boldsymbol{e}}_k^{\rm{H}}} \right|}^2}, \forall k.} \label{eq.Zk}
\end{align}
In fact, the considered LIS system is significantly different from a classical massive MIMO because of a \emph{key difference in the SINR expression}.
In the considered LIS system, the desired signal power, ${{\left| {{{\boldsymbol{h}}_{kk}}} \right|}^4}$, is calculated by the squared sum of the squared LOS channel gains over all antennas, i.e., $\big(\sum\nolimits_{m=1}^M({\beta _{kkm}^{\rm{L}}})^2\big)^2$,
and this is a deterministic value known at the LIS by measuring the signal strength of the reference signals.
However, in a conventional massive MIMO system, this desired signal power, ${{\left| {{{\boldsymbol{h}}_{kk}}} \right|}^4}$, is not a deterministic value and cannot be known at the BS accurately because of an NLOS fading.
Therefore, the BS can detect the desired signal using only the estimated CSI, ${{ {{{\boldsymbol{\hat h}}_{kk}}} }}$,
resulting in $S_k = {{| {{{\boldsymbol{\hat h}}_{kk}}} |}^4}$ and ${X_k} = {{| {{\boldsymbol{\hat h}}_{kk}^{\rm{H}}{{\boldsymbol{e}}_{k}}} |}^2}$
as most prior studies on massive MIMO systems have considered (e.g., see \cite{ref.How, ref.Mine}, and references therein).
Given the uplink data rate $R_k$, we will analyze the moments of mutual information asymptotically as $M$ and $K$ increase without bounds.

\section{Asymptotic Rate Analysis}
We consider an LIS-based large antenna-array system composed of a large number of discrete antennas that are densely distributed on a contiguous LIS and each LIS unit occupies a subarea of the LIS with $M$ antennas.
Given a massive number of IoT devices will be connected via wireless communication systems in the near future \cite{ref.IoT},
we present an asymptotic analysis of the data rate in an LIS-based system as $M$ and $K$ increase. 
In conventional massive MIMO systems, there is a relationship between $M$ and $K$ such that $M/K \ge {\rm{1}}$ and $M/K$ is constant as in \cite{ref.How} and \cite{ref.Scaling}. 
In contrast, LIS enables wireless communications without any constraint on the relationship between $M$ and $K$. 

From (\ref{eq.Sk_def}), we can describe the desired signal power, ${S_k}$, as 
\begin{equation}
{S_k} = {\left| {{{\boldsymbol{h}}_{kk}}} \right|^4} = {\left( {\sum\limits_m^M {{{\left| {\beta _{km}^{\rm{L}}{h_{kkm}}} \right|}^2}} } \right)^2} = {\left( {\sum\limits_m^M {{{\left( {\beta _{km}^{\rm{L}}} \right)}^2}} } \right)^2}. 
\end{equation}
$\sum\nolimits_m^M {{{\left( {\beta _{km}^{\rm{L}}} \right)}^2}}$ is the summation of the desired signal power received at LIS unit $k$,
and this is equivalent to the summation of the power received within the ranges of $-L \le x \le L$ and $-L \le y \le L$ when the signal is transmitted from the location of $(0,0,{z_k})$ in Cartesian coordinates \cite{ref.LIS1}. 
Then, ${S_k}$ converges as ${S_k} - {{{{\bar S}_k}}^2} \xrightarrow[M \to \infty ]{} 0$                         
where we define
\begin{align}
{{\bar S}_k} &= {\frac{1}{{4\pi \Delta {L^2}}}}\iint_{{ - L \le \left( {x,y} \right) \le L}} {\frac{{{z_k}}}{{{{\left( {x^2 + y^2 + z_k^2} \right)}^{\frac{3}{2}}}}}} {\rm{d}}x{\rm{d}}y \nonumber \\
&{={\frac{1}{{2\pi \Delta {L^2}}}}\int_{{ - L \le  {y}  \le L}} {\frac{{{z_k}L}}{{{{\left( {y^2 + z_k^2} \right)\sqrt{L^2+y^2+z_k^2}}}}}} {\rm{d}}y} \nonumber \\
&= {\frac{1}{{\pi \Delta {L^2}}}}{\tan ^{ - 1}}\left( {\frac{{{L^2}}}{{{z_k}\sqrt {2{L^2} + z_k^2} }}} \right)  = {\frac{{p_k}}{{\pi \Delta {L^2}}}}, \forall k  \label{eq.S_bar_k}
\end{align}
where ${p_k} = {\tan ^{ - 1}}\left( {\frac{{{L^2}}}{{{z_k}\sqrt {2{L^2} + z_k^2} }}} \right)$. 
Then, we have
\begin{equation}
{S_k} - {\bar p_k} \xrightarrow[M \to \infty ]{} 0, \label{eq.sk}
\end{equation}
where ${\bar p_k} = \frac{{p_k^2}}{{{\pi ^2}\Delta {L^4}}}$. 
(\ref{eq.sk}) shows that ${S_k}$ converges to a constant value, ${\bar p_k}$, depending on ${z_k}$ and $L$,
and the total captured energy by the LIS can increase as $M$ increases within the constrained physical area of LIS unit (i.e., $\Delta L^2$ decreases).
In fact, the work in \cite{ref.Poon2005Dof} showed that the total captured energy is limited by the product of the physical area used for deploying antennas and the channel's solid angle, 
and it remains unchanged as $M$ increases within the constrained physical area.
However, since the authors in \cite{ref.Poon2005Dof} assumed perfect NLOS channels
and obtained the array response by using the far-field approximation, 
\cite{ref.Poon2005Dof} does not directly apply to our LIS system.
In the considered LIS system, the increase of $M$ indicates an increase of the number of LOS paths and this results in an increase in the number of spatial channels and the channel's solid angle.
Then, the total captured energy by an LIS can increase as $M$ increases within the constrained physical area of the LIS unit.
Therefore, the mean and variance of ${R_k}$ can be readily derived, as follows. 

{\bf{{Corollary 1.}}} The mean and variance of ${R_k}$ can be respectively approximated as follows:
\begin{align}
{\mu _{{R_k}}} &\approx \log \left( {1 + {\mu _{{\gamma _k}}}} \right) - \frac{{\sigma _{{\gamma _k}}^2}}{{2{{\left( {1 + {\mu _{{\gamma _k}}}} \right)}^2}}}, \label{eq.MRk0}\\
\sigma _{{R_k}}^2 &\approx \frac{{\sigma _{{\gamma _k}}^2}}{{{{\left( {1 + {\mu _{{\gamma _k}}}} \right)}^2}}} - \frac{{\sigma _{{\gamma _k}}^4}}{{4{{\left( {1 + {\mu _{{\gamma _k}}}} \right)}^4}}}, \label{eq.SRk0}
\end{align}
where ${\mu _{{\gamma _k}}}$ and $\sigma _{{\gamma _k}}^2$ represent the mean and variance of the uplink SINR, respectively, which are likewise approximated as
\begin{align}
{\mu _{{\gamma _k}}} &\approx {\rho _k}{S_k}\left( {1 - \tau _k^2} \right)\left( {\frac{1}{{{\mu _{{I_k}}}}} + \frac{{\sigma _{{I_k}}^2}}{{\mu _{{I_k}}^3}}} \right), \label{eq.MSINRk}\\
\sigma _{{\gamma _k}}^2 &\approx \rho _k^2S_k^2{\left( {1 - \tau _k^2} \right)^2}\left( {\frac{{\sigma _{{I_k}}^2}}{{\mu _{{I_k}}^4}} - \frac{{\sigma _{{I_k}}^4}}{{\mu _{{I_k}}^6}}} \right), \label{eq.SSINRk}
\end{align}
\begin{proof}
From \cite{ref.Taylor}, the mean of a function $f$ for a random variable $X$ using Taylor expansions can be approximated by 
\begin{equation}
{\rm{E}}\left[ {f\left( X \right)} \right] \approx f\left( {{\mu _X}} \right) + \frac{{f''\left( {{\mu _X}} \right)}}{2}\sigma _X^2. 
\end{equation}
The variance of a function $f$ for a random variable $X$ can be approximated as
\begin{equation}
{\rm{Var}}\left[ {f\left( X \right)} \right] \approx {\left( {f'\left( {{\mu _X}} \right)} \right)^2}\sigma _X^2 - \frac{{{{\left( {f''\left( {{\mu _X}} \right)} \right)}^2}\sigma _X^4}}{4}. 
\end{equation}
Since $S_k$ is constant, (\ref{eq.MSINRk}) and (\ref{eq.SSINRk}) can be obtained when $f\left( {{I_k}} \right) = {\rho _k}{S_k}\left( {1 - \tau _k^2} \right)/{I_k}$ from (\ref{eq.simpleSINR2}).
Similarly, (\ref{eq.MRk0}) and (\ref{eq.SRk0}) can be obtained when $f\left( {{\gamma _k}} \right) = \log \left( {1 + {\gamma _k}} \right)$.
\end{proof}

Corollary 1 shows that both the mean and variance of the uplink rate are determined exclusively by a random variable $I_k$. 
Based on the results from Corollary 1, the mean and variance of $I_k$ will be analyzed asymptotically.

\subsection{Asymptotic Analysis of $R_k$}\label{section.Analysis}
We provide an asymptotic analysis of uplink data rate, $R_k$, by following three steps.
Given that $R_k$ exclusively depends on a random variable $I_k$, we first analyze the moments of random variables $X_k$, $Y_{jk}$, and $Z_k$ from (\ref{eq.Ik_def}).
We then asymptotically obtain the asymptotic moments of $I_k$ given the covariances between $X_k$, $Y_{jk}$, and $Z_k$.
We finally derive asymptotic moments of $R_k$ from Corollary 1 using the derived asymptotic moments of $I_k$.

In order to obtain the moments of $I_k$, we first derive the following lemmas from the asymptotic analyses.

{\bf{{Lemma 1.}}} The mean and variance of $X_k$ follow ${\mu _{{X_k}}} = \sum\nolimits_m^M {{{\left( {\beta _{km}^{\rm{L}}} \right)}^4}}$ and 
$\sigma _{{X_k}}^2 = {\left( {\sum\nolimits_m^M {{{\left( {\beta _{km}^{\rm{L}}} \right)}^4}} } \right)^2}$, respectively.
\begin{proof}
The detailed proof is presented in Appendix A.
\end{proof}


{\bf{{Lemma 2.}}} The mean and variance of $Y_{jk}$ follow ${\mu _{{Y_{jk}}}} - {\bar \mu _{{Y_{jk}}}}\xrightarrow[M \to \infty ]{} 0$ and 
$\sigma _{{Y_{jk}}}^2 - \bar \sigma _{{Y_{jk}}}^2 \xrightarrow[M \to \infty ]{} 0$, respectively, where 
\begin{align}
{\bar \mu _{{Y_{jk}}}} &= s_{jk}^{\rm{L}} + s_{jk}^{\rm{N1}} + s_{jk}^{{\rm{N2}}} + {\left| {{\mu _{jk}^{\rm{L}}}} \right|^2}, \label{req.B.MY}\\
\bar \sigma _{{Y_{jk}}}^2 &= {\left( {s_{jk}^{\rm{L}} + s_{jk}^{\rm{N1}} + s_{jk}^{{\rm{N2}}}} \right)^2} + 2{\left| {{\mu _{jk}^{\rm{L}}}} \right|^2}\left( {s_{jk}^{\rm{L}} + s_{jk}^{\rm{N1}} + s_{jk}^{{\rm{N2}}}} \right), \label{req.B.SY}
\end{align}
and
\begin{align}
{\mu _{jk}^{\rm{L}}} &= \sqrt {\frac{{{\kappa _{jk}}\left( {1 - \tau _k^2} \right)}}{{{\kappa _{jk}} + 1}}} {\boldsymbol{h}}_{kk}^{\rm{H}}{\boldsymbol{h}}_{jk}^{\rm{L}}, \label{req.B.ML} \\
s_{jk}^{\rm{L}} &= \frac{{{\kappa _{jk}}\tau _k^2}}{{{\kappa _{jk}} + 1}}\sum\limits_m^M {{{\left( {\beta _{km}^{\rm{L}}\beta _{jm}^{\rm{L}}} \right)}^2}} , \label{req.B.SL} \\
s_{jk}^{\rm{N1}} &= \frac{{1 - \tau _k^2}}{{{\kappa _{jk}} + 1}}{\sum\limits_p^P {\left| {{\boldsymbol{h}}_{kk}^{\rm{H}}{{\boldsymbol{r}}_{jkp}}} \right|} ^2}, \label{req.B.SN1}\\
s_{jk}^{{\rm{N2}}} &= \frac{{\tau _k^2}}{{{\kappa _{jk}} + 1}}\sum\limits_{m,p}^{M,P} {{{\left( {\alpha _{jkp}^{{\rm{NL}}}\beta _{km}^{\rm{L}}l_{jkm}^{{\rm{NL}}}} \right)}^2}} /M. \label{req.B.SN2}
\end{align}\vspace{-20pt}
\begin{proof}
The detailed proof is presented in Appendix B.
\end{proof}

{\bf{{Lemma 3.}}} The mean and variance of $Z_k$ follow 
${\mu _{{Z_{k}}}} - {\bar \mu _{{Z_{k}}}}\xrightarrow[M \to \infty ]{} 0$ and 
$\sigma _{{Z_{k}}}^2 - \bar \sigma _{{Z_{k}}}^2 \xrightarrow[M \to \infty ]{} 0$, respectively, where
${\bar \mu _{{Z_{k}}}} = \sum\nolimits_m^M {{{\left( {\beta _{km}^{\rm{L}}} \right)}^2}} + \mu_{Z_k^{\rm{w}}}$,
$\bar \sigma _{{Z_{k}}}^2 = \tau _k^2\left( {2 - \tau _k^2} \right)\sum\nolimits_m^M {{{\left( {\beta _{km}^{\rm{L}}} \right)}^4}} + \sigma_{Z_k^{\rm{w}}}^2$,
and
\begin{align}
\mu_{Z_k^{\rm{w}}}&=\sum\limits_m^M \left(\sigma_{{z_{km}^{\rm{wL}}}}^2 + \sigma_{{z_{km}^{\rm{wR}}}}^2 +
2 {\rm{Re}}\big( \omega_{km}^{\rm{wLR}} \big)\right),\label{req.C.Mzw}\\
\sigma_{Z_k^{\rm{w}}}^2&= \left(\sum\limits_m^M \left(\sigma_{{z_{km}^{\rm{wL}}}}^2 + \sigma_{{z_{km}^{\rm{wR}}}}^2 + 2 {\rm{Re}}\big( \omega_{km}^{\rm{wLR}} \big)\right)\right)^2.\label{req.C.Vzw}
\end{align}
The terms $\sigma_{{z_{km}^{\rm{wL}}}}^2$, $\sigma_{{z_{km}^{\rm{wR}}}}^2$, and $ \omega_{km}^{\rm{wLR}}$ are given in (\ref{eq.C.SwL}), (\ref{eq.C.SwR}), and (\ref{eq.omega_km}), respectively.
\begin{proof}
The detailed proof is presented in Appendix C.
\end{proof}

$\sum\nolimits_m^M {{{\left( {\beta _{km}^{\rm{L}}} \right)}^2}} $ in Lemma 3 can be asymptotically obtained by $\sqrt{\bar p_k}$ using (\ref{eq.sk}). 
Similarly, $\sum\nolimits_m^M {{{\left( {\beta _{km}^{\rm{L}}} \right)}^4}} $ in Lemma 1 and 3 can be obtained by 
$\sum\nolimits_m^M {{{\left( {\beta _{km}^{\rm{L}}} \right)}^4}}  - {\bar q_k} \xrightarrow[M \to \infty ]{} 0$, 
where ${\bar q_k} = \frac{{{q_k}}}{{16{\pi ^2}\Delta {L^2}}}$ and ${q_k}$ is given by
\begin{align}
{q_k} &= \iint_{{ - L \le \left( {x,y} \right) \le L}} {\frac{{z_k^2}}{{{{\left( {{x^2} + {y^2} + z_k^2} \right)}^3}}}} {\rm{d}}x{\rm{d}}y\nonumber\\
&= \frac{{{L^2}}}{{\left( {{L^2} + z_k^2} \right)\left( {2{L^2} + z_k^2} \right)}} + \frac{{L\left( {2{L^2} + 3z_k^2} \right)}}{{{{\left( {z_k^2\left( {{L^2} + z_k^2} \right)} \right)}^{3/2}}}}{\tan ^{ - 1}}\left( {\frac{L}{{\sqrt {{L^2} + z_k^2} }}} \right). 
\end{align}

Next, we asymptotically derive the covariances between $X_k$, $Y_{jk}$, and $Z_k$, and then the asymptotic mean and variance of $I_k$ are obtained by the following lemma. 

{\bf{{Lemma 4.}}} The mean and variance of $I_k$ follow ${\mu _{{I_k}}} - {\bar \mu _{{I_k}}}\xrightarrow[M \to \infty ]{} 0$ and $\sigma _{{I_k}}^2/M^2 - \bar \sigma _{{I_k}}^2/M^2\xrightarrow[M,K \to \infty ]{} 0$, respectively,
where $\bar{\mu _{{I_k}}}$ and $\bar\sigma _{{I_k}}^2$ are given as
\begin{align}
{\bar \mu _{{I_k}}} &= {\rho _k}\tau _k^2{\bar q_k} + \sum\limits_{j \ne k}^K {{\rho _j}{{\bar \mu }_{{Y_{jk}}}}}  + \sqrt {{{\bar p}_k}} {+ \mu_{Z_k^{\rm{w}}}},\\
\bar \sigma _{{I_k}}^2 &= \rho _k^2\tau _k^4\bar q_k^2 + \tau _k^2\left( {2 - \tau _k^2} \right){\bar q_k} + \sum\limits_{j \ne k}^K {\rho _j^2\bar \sigma _{{Y_{jk}}}^2}  + \sum\limits_{{i,j\ne k}:{i\ne j}}^K {\rho _i\rho _j{{\bar \omega }_{ijk}}}
{+\sigma_{Z_k^{\rm{w}}}^2}, 
\end{align}
where
\begin{align}
{\bar\omega _{ijk}} &= 2{\mathop{\rm Re}\nolimits} \left( {{\mu _{{c_{ik}}}}{\mu _{c_{jk}}^*}\sum\nolimits_m^M {{\mu _{a_{imk}}^*}{\mu _{{a_{jmk}}}}} } \right), \label{req.D.Cov}\\
{\mu _{{c_{tk}}}} &= \sqrt {\frac{{{\kappa _{tk}}\left( {1 - \tau _k^2} \right)}}{{{\kappa _{tk}} + 1}}} {\boldsymbol{h}}_{kk}^{\rm{H}}{\boldsymbol{h}}_{tk}^{\rm{L}},\\
{\mu _{{a_{tmk}}}} &= \sqrt {\frac{{\tau _k^2{\kappa _{tk}}}}{{{\kappa _{tk}} + 1}}} \beta _{km}^{\rm{L}}\beta _{tm}^{\rm{L}}{h_{tkm}}.
\end{align}\vspace{-20pt}
\begin{proof}
The detailed proof is presented in Appendix D.
\end{proof}

Lemma 4 shows that ${\bar \mu _{{I_k}}}$ and $\bar \sigma _{{I_k}}^2$ are deterministic values depending on locations of the devices and the correlation matrices. 
Therefore, we can approximate the mean and variance of $R_k$ as follows.

{\bf{Theorem 1.}} The mean and variance of $R_k$ follow ${\mu _{{R_k}}} - {\bar \mu _{{R_k}}}\xrightarrow[M \to \infty ]{} 0$ and $\sigma _{{R_k}}^2 - \bar \sigma _{{R_k}}^2\xrightarrow[M,K \to \infty ]{} 0$,
where ${\bar \mu _{{R_k}}}$ and $\bar \sigma _{{R_k}}^2$ are
\begin{align}
{\bar \mu _{{R_k}}} &= \log \left( {1 + {{\bar \mu }_{{\gamma _k}}}} \right) - \frac{{\bar \sigma _{{\gamma _k}}^2}}{{2{{\left( {1 + {{\bar \mu }_{{\gamma _k}}}} \right)}^2}}}, \label{eq.MRk}\\
\bar \sigma _{{R_k}}^2 &= \frac{{\bar \sigma _{{\gamma _k}}^2}}{{{{\left( {1 + {{\bar \mu }_{{\gamma _k}}}} \right)}^2}}} - \frac{{\bar \sigma _{{\gamma _k}}^4}}{{4{{\left( {1 + {{\bar \mu }_{{\gamma _k}}}} \right)}^4}}}. \label{eq.SRk}
\end{align}
\begin{proof}
By respectively replacing ${\mu _{{I_k}}}$ and $\sigma _{{I_k}}^2$ in (\ref{eq.MSINRk}) and (\ref{eq.SSINRk}) with ${\bar \mu _{{I_k}}}$ and $\bar \sigma _{{I_k}}^2$ from Lemma 4, 
we obtain the mean and variance of the asymptotic uplink SINR.
Let us define the mean and variance of the asymptotic uplink SINR of device $k$ as ${\bar \mu _{{\gamma _k}}}$ and $\bar \sigma _{{\gamma _k}}^2$, respectively,
then ${\bar \mu _{{\gamma _k}}}$ and $\bar \sigma _{{\gamma _k}}^2$ are also deterministic values, 
as follows:
\begin{align}
{\bar\mu _{{\gamma _k}}} &= {\rho _k}{{\bar p_k}}\left( {1 - \tau _k^2} \right)\left( {\frac{1}{{{\bar\mu _{{I_k}}}}} + \frac{{\bar\sigma _{{I_k}}^2}}{{\bar\mu _{{I_k}}^3}}} \right), \\
\bar\sigma _{{\gamma _k}}^2 &= \rho _k^2{{\bar p_k^2}}{\left( {1 - \tau _k^2} \right)^2}\left( {\frac{{\bar\sigma _{{I_k}}^2}}{{\bar\mu _{{I_k}}^4}} - \frac{{\bar\sigma _{{I_k}}^4}}{{\bar\mu _{{I_k}}^6}}} \right).
\end{align}
We can obtain $
{\bar \mu _{{R_k}}}$ and $\bar \sigma _{{R_k}}^2$ by respectively replacing ${\mu _{{\gamma_k}}}$ and $\sigma _{{\gamma_k}}^2$ in (\ref{eq.MRk0}) and (\ref{eq.SRk0}) with ${\bar\mu _{{\gamma_k}}}$ and $\bar\sigma _{{\gamma_k}}^2$, which completes the proof.
\end{proof}
We refer to ${\bar \mu _{{R_k}}}$ and $\bar \sigma _{{R_k}}^2$ as the asymptotic mean and variance of $R_k$, respectively.
Given deterministic values of $\bar \mu _{\gamma_k}$ and $\bar \sigma _{\gamma_k}^2$, Theorem 1 shows that the mean and variance of uplink data rate in an LIS-based large antenna-array system can be obtained based on deterministic values such as the locations of the devices and the correlation matrices.
Then, we can evaluate the performance of an LIS-based system in terms of ergodic rate, reliability, and scheduling diversity, without extensive simulations.
In particular, we can easily estimate the ergodic rate from (\ref{eq.MRk}), and verify system reliability and the scheduling diversity gain from (\ref{eq.SRk}). 
Furthermore, the results from Theorem 1 will be in close agreement with the moments of mutual information resulting from an actual LIS-based system as the number of devices and antennas increase.



\section{Channel Hardening Effect and Performance Bound}
The channel hardening effect is an important feature in large antenna-array systems whereby the variance of mutual information shrinks as the number of antennas grows \cite{ref.Hard}.
Since a wireless system's reliability and scheduling diversity depend on the fluctuations of the mutual information, 
it is important to estimate the fluctuations that can be expected in a given large antenna-array systems such as an LIS.
Given the importance of the channel hardening effect, next, we verify its occurrence in an LIS-based large antenna-array system and we then derive the performance bound of the ergodic rate. 

Since $M = {\left( {2L/\Delta L} \right)^2}$, (\ref{eq.simpleSINR2}) is given by using (\ref{eq.sk}) as
\begin{equation}
{\bar \gamma _k} = \frac{{{\rho _k}\left( {1 - \tau _k^2} \right)\frac{{p_k^2}}{{{L^4}{\pi ^2}}}}}{{16{{\bar I}_k}/{M^2}}}, \label{eq.SINRbar}
\end{equation}
where ${\bar \gamma _k}$ denotes the asymptotic value of ${\gamma _k}$ and ${\bar I_k}$ denotes a random variable with a mean and variance of ${\bar \mu _{{I_k}}}$ and $\bar \sigma _{{I_k}}^2$ from Lemma 4, respectively.
With $M = {\left( {2L/\Delta L} \right)^2}$, ${\bar \mu _{{I_k}}}$ and $\bar \sigma _{{I_k}}^2$ are represented by
\begin{align}
{\bar \mu _{{I_k}}} &= {\frac{{M{\rho _k}\tau _k^2{q_k}}}{{64{\pi ^2}{L^2}}} + \frac{{M{p_k}}}{{4\pi {L^2}}}} + \sum\limits_{j \ne k}^K {{\rho _j}{{\bar \mu }_{{Y_{jk}}}}}{+\mu_{Z_k^{\rm{w}}}}  ,\label{eq.MIk}\\
\bar \sigma _{{I_k}}^2 &= {\frac{{{M^2}\rho _k^2\tau _k^4q_k^2}}{{4096{\pi ^4}{L^4}}} + \frac{{M\tau _k^2{q_k}\left( {2 - \tau _k^2} \right)}}{{64{\pi ^2}{L^2}}} + \sum\limits_{j \ne k}^K {\rho _j^2\bar \sigma _{{Y_{jk}}}^2}  + \sum\limits_{{i,j\ne k}:{i\ne j}}^K {\rho _i\rho _j{{\bar \omega }_{ijk}}} } {+\sigma_{Z_k^{\rm{w}}}^2}, \label{eq.SIk}
\end{align}
We can observe in (\ref{eq.SINRbar}) that the mean and variance of ${\bar \gamma _k}$ are determined by ${\bar \mu _{{I_k}}}/{M^2}$ and $\bar \sigma _{{I_k}}^2/{M^4}$, respectively.
Lemma 5 is used to determine the scaling laws of ${\bar \mu _{{I_k}}}/{M^2}$ and $\bar \sigma _{{I_k}}^2/{M^4}$ according to $M$. 

{\bf{{Lemma 5.}}} According to the scaling laws for $M$, the mean and variance of ${\bar I_k}/{M^2}$ follow ${\bar \mu _{{I_k}}}/{M^2} - {\hat \mu _{{I_k}}}\xrightarrow[M \to \infty ]{} 0$ and $\bar \sigma _{{I_k}}^2/{M^4}\xrightarrow[M \to \infty ]{} 0$, respectively, where
${\hat \mu _{{I_k}}} {=} \sum\nolimits_{j \ne k}^K {\frac{{{\rho _j}{\kappa _{jk}}\left( {1 - \tau _k^2} \right)}}{{{M^2}\left( {1+{\kappa _{jk}}} \right)}}{{\left| {{\boldsymbol{h}}_{kk}^{\rm{H}}{\boldsymbol{h}}_{jk}^{\rm{L}}} \right|}^2}}$. 
\begin{proof}
From (\ref{eq.SIk}), ${{\bar \sigma _{{I_k}}^2}}/{{{M^4}}}$ is obtained as follows:
\begin{equation}
\frac{{\bar \sigma _{{I_k}}^2}}{{{M^4}}} = \frac{{\rho _k^2\tau _k^4q_k^2}}{{4096{\pi ^4}{L^4}{M^2}}} + \frac{{\tau _k^2{q_k}\left( {2 - \tau _k^2} \right)}}{{64{\pi ^2}{L^2}{M^3}}}{+\frac{\sigma_{Z_k^{\rm{w}}}^2}{M^4}}+ \frac{{\sum\nolimits_{j \ne k}^K {\rho _j^2\bar \sigma _{{Y_{jk}}}^2}  + \sum\nolimits_{{i,j\ne k}:{i\ne j}}^K {\rho _i\rho _j{{\bar \omega }_{ijk}}} }}{{{M^4}}}. \label{eq.SIM4}
\end{equation}
From the scaling laws for $M$, as discussed in{\cite{ref.Mine}}, $\frac{{\rho _k^2\tau _k^4q_k^2}}{{4096{\pi ^4}{L^4}{M^2}}}$ and $\frac{{\tau _k^2{q_k}\left( {2 - \tau _k^2} \right)}}{{64{\pi ^2}{L^2}{M^3}}}$ in (\ref{eq.SIM4}) converge to zero as $M$ goes to infinity.
${\sigma_{Z_k^{\rm{w}}}^2}$ in (\ref{eq.SIM4}) is calculated by the squared sum of $M$ elements from (\ref{req.C.Vzw}),
and then ${\sigma_{Z_k^{\rm{w}}}^2}$ increases with $\mathcal{O}({M^2})$ as $M$ increases based on the scaling law.
Hence, ${\sigma_{Z_k^{\rm{w}}}^2}/M^4$ converges to zero as $M \to \infty$.
Also,
${{\bar \sigma _{{Y_{jk}}}^2}}/{{{M^4}}}$ in (\ref{eq.SIM4}) is given by using (\ref{req.B.SY}) as follows:
\begin{equation}
\frac{{\bar \sigma _{{Y_{jk}}}^2}}{{{M^4}}} = {\left( {\frac{{s_{jk}^{\rm{L}} + s_{jk}^{\rm{N1}} + s_{jk}^{{\rm{N2}}}}}{{{M^2}}}} \right)^2} + \frac{{2{{\left| {{\mu _{jk}^{\rm{L}}}} \right|}^2}\left( {s_{jk}^{\rm{L}} + s_{jk}^{\rm{N1}} + s_{jk}^{{\rm{N2}}}} \right)}}{{{M^4}}}. \label{eq.SYM4}
\end{equation}
In order to verify the scaling law of (\ref{eq.SYM4}), we determine the scaling laws of ${s_{jk}^{\rm{L}}}$, ${s_{jk}^{\rm{N1}}}$, ${s_{jk}^{\rm{N2}}}$, and ${{{\left| {{\mu _{jk}^{\rm{L}}}} \right|}^2}}$ according to $M$.
From (\ref{req.B.SL}), ${s_{jk}^{\rm{L}}}$ is calculated by the sum of ${\left( {\beta _{km}^{\rm{L}}\beta _{jm}^{\rm{L}}} \right)^2}$ over all $m$ where $m = 1, \ldots ,M$. 
Consequently, ${s_{jk}^{\rm{L}}}$ increases with $\mathcal{O}(M)$ as $M$ increases.
From (\ref{req.B.SN1}), ${s_{jk}^{\rm{N1}}}$ is calculated by the sum of $\left| {{\boldsymbol{h}}_{kk}^{\rm{H}}{{\boldsymbol{r}}_{jkp}}} \right|^2$ over all $p$ where $p = 1, \ldots ,P$. 
Given that ${{\boldsymbol{r}}_{jkp}}$ is a correlation vector normalized by $\sqrt{M}$ from (\ref{eq.DNLOS}) and
${{\boldsymbol{h}}_{kk}^{\rm{H}}{{\boldsymbol{r}}_{jkp}}}$ is calculated by the sum of $M$ elements,
${s_{jk}^{\rm{N1}}}$ increases with $\mathcal{O}({M})$ as $M$ increases.
From (\ref{req.B.SN2}), ${s_{jk}^{\rm{N2}}}$ is calculated by the sum of ${{{\left( {\alpha _{jkp}^{{\rm{NL}}}\beta _{km}^{\rm{L}}l_{jkm}^{{\rm{NL}}}} \right)}^2}}/M$ for all $m$ and $p$.
Thus, ${s_{jk}^{\rm{N2}}}$ follows $\mathcal{O}(1)$ as $M$ increases.
From (\ref{req.B.ML}), ${{{\left| {{\mu _{jk}^{\rm{L}}}} \right|}}}$ is obtained from ${\boldsymbol{h}}_{kk}^{\rm{H}}{\boldsymbol{h}}_{jk}^{\rm{L}}$ which is calculated by the sum of $M$ elements.
Therefore ${{{\left| {\mu _{jk}^{\rm{L}}} \right|}}}^2$ increases with $\mathcal{O}({M}^2)$ as $M$ increases.
Hence, (\ref{eq.SYM4}) goes to zero as $M \to \infty$ and we have $\sum\nolimits_{j \ne k}^K {\rho _j^2\bar \sigma _{{Y_{jk}}}^2} /{M^4}\xrightarrow[M \to \infty ]{}0$\footnote{ If the number of devices that dominantly transmits interference signals to a target LIS unit is as large as $M$, it will not converge to 0. 
However, there are actually far fewer such devices than $M$ because the minimum distance of the $(x,y)$ coordinates between adjacent devices is considered under the assumption that the LIS units do not overlap.
Therefore, we assume that this value always converges to 0. \vspace{-0.5cm}}.
Similarly, ${\bar \omega _{ij}}$ from (\ref{req.D.Cov}) increases with $\mathcal{O}({M}^3)$ as $M$ increases.
Then, $\sum_{{i,j\ne k}:{i\ne j}}^K {\rho _i\rho _j{{\bar \omega }_{ijk}}} /{M^4}\xrightarrow[M \to \infty ]{}0$
and (\ref{eq.SIM4}) eventually converges to zero as $M\to\infty$.

From (\ref{eq.MIk}), ${{{\bar \mu }_{{I_k}}}/{M^2}}$ is obtained by
\begin{equation}
\frac{{{{\bar \mu }_{{I_k}}}}}{{{M^2}}} = \frac{{{\rho _k}\tau _k^2{q_k}}}{{64{\pi ^2}{L^2}M}}+ \frac{{{p_k}}}{{4\pi {L^2}M}} {+\frac{\sigma_{Z_k^{\rm{w}}}^2}{M^2}}+ \sum\limits_{j \ne k}^K \frac{{\rho _j}{{\bar \mu }_{{Y_{jk}}}}}{M^2} .\label{eq.MIM2}
\end{equation}
From the scaling laws for $M$, $\frac{{{\rho _k}\tau _k^2{q_k}}}{{64{\pi ^2}{L^2}M}}$ and $\frac{{{p_k}}}{{4\pi {L^2}M}}$ in (\ref{eq.MIM2}) converge to zero as $M\to\infty$.
${\mu_{Z_k^{\rm{w}}}^2}$ in (\ref{eq.MIM2}) is calculated by the sum of $M$ elements from (\ref{req.C.Mzw}),
and then ${\mu_{Z_k^{\rm{w}}}}$ increases with $\mathcal{O}({M})$, 
finally ${\mu_{Z_k^{\rm{w}}}^2}/M^2$ converges to zero as $M \to \infty$.
Also, ${{{\bar \mu }_{{Y_{jk}}}}/{M^2}}$ in (\ref{eq.MIM2}) is presented by using (\ref{req.B.MY}) as
$\frac{{{{\bar \mu }_{{Y_{jk}}}}}}{{{M^2}}} = \frac{{s_{jk}^{\rm{L}} + s_{jk}^{\rm{N1}} + s_{jk}^{{\rm{N2}}} + {{\left| {{\mu _{jk}^{\rm{L}}}} \right|}^2}}}{{{M^2}}}. $
From the scaling laws of ${s_{jk}^{\rm{L}}}$, ${s_{jk}^{\rm{N1}}}$, ${s_{jk}^{\rm{N2}}}$, and ${{{\left| {{\mu _{jk}^{\rm{L}}}} \right|}^2}}$ according to $M$, we have 
${{{{\bar \mu }_{{Y_{jk}}}}}}/{{{M^2}}} - {{{{\left| {{\mu _{jk}^{\rm{L}}}} \right|}^2}}}/{{{M^2}}}\xrightarrow[M \to \infty ]{}0$.
Therefore, (\ref{eq.MIM2}) is asymptotically obtained as
$\frac{{{{\bar \mu }_{{I_k}}}}}{{{M^2}}}- \sum\limits_{j \ne k}^K \frac{{\rho _j}{{{\left| {{\mu _{jk}^{\rm{L}}}} \right|}^2}}}{M^2} \xrightarrow[M \to \infty ]{}0,$
which completes the proof.
\end{proof}

Lemma 5 shows that ${\bar I _k}/M^2$ does not converge to a random variable, but rather to a constant without any variance as $M$ increases.
Therefore, we can prove the following result related to the occurrence of the channel hardening effect and the performance bound of the uplink data rate.

{\bf{{Theorem 2.}}} The asymptotic variance of $R_k$ goes to zero as $M\to\infty$, and an asymptotic mean of $R_k$ is given by
${\bar \mu _{{R_k}}} - {\hat \mu _{{R_k}}} \xrightarrow[M \to \infty ]{} 0$, 
where ${\hat \mu _{{R_k}}} = \log \left( {1 + \frac{{p_k^2{\rho _k}\left( {1 - \tau _k^2} \right)}}{{16{L^4}{\pi ^2}{{\hat \mu }_{{I_k}}}}}} \right)$
is an asymptotic bound of the uplink data rate.

\begin{proof}
From Lemma 5, ${\bar I _k}/M^2$ converges on a constant value of ${{\hat \mu }_{{I_k}}}$ as $M \to \infty$.
Then, ${\bar \gamma _k}$ from (\ref{eq.SINRbar}) also converges on a constant value without any variance as $M \to \infty$, as given by
${\bar \gamma _k}- \frac{{p_k^2{\rho _k}( {1 - \tau _k^2} )}}{{16{L^4}{\pi ^2}{{\hat \mu }_{{I_k}}}}}\xrightarrow[M \to \infty ]{}0.$
Therefore, the asymptotic uplink data rate, $\bar R _k$, converges in distribution to a constant value:
$\bar R _k  - \log \Big( 1 +  \frac{{p_k^2{\rho _k}( {1 - \tau _k^2} )}}{{16{L^4}{\pi ^2}{{\hat \mu }_{{I_k}}}}} \Big)\xrightarrow[M \to \infty ]{} 0.$
Finally, the asymptotic mean and variance of $R_k$ converge, respectively, as
${\bar\mu _{{R_k}}} - \log \left( 1 +  \frac{{p_k^2{\rho _k}( {1 - \tau _k^2} )}}{{16{L^4}{\pi ^2}{{\hat \mu }_{{I_k}}}}} \right)\xrightarrow[M \to \infty ]{}0$ and
$\bar\sigma _{{R_k}}^2\xrightarrow[M,K \to \infty ]{}0.$
\end{proof}

\begin{table}[!t]
\renewcommand{\arraystretch}{1.3}
\caption{Simulation parameters}
\label{table.para}
\centering
\begin{tabular}{|l|c|}
\hline
\textbf{Parameter} & \textbf{Value}\\
\hline
Carrier frequency & $3$ GHz \\
\hline
Uplink target SNR & $3$ dB \\
\hline
{Hardware impairments $(\delta)$} & {$1$} \\
\hline
Channel imperfectness $(\tau_k^2)$ & $0.5$ \\
\hline
Length of LIS unit $(2L)$ & $0.5$ $\rm{m}$ \\
\hline
Rician factor $({\kappa}[${dB}$])$ \cite{ref.LTE} & $13 - 0.03{d[\rm{m}]}$ \\
\hline
LOS path loss model \cite{ref.TSE}& $11 +20{\rm{log}}_{10}d[\rm{m}]$ \\
\hline
NLOS path loss model \cite{ref.How} &  $37{\rm{log}}_{10}d[\rm{m}]$ $(\beta_{\rm{PL}}=3.7)$\\
\hline
\end{tabular}
\end{table}

Theorem 2 shows that the channel fading of an LIS-based large antenna-array system behaves as a static channel and its impact on the uplink data rate becomes negligible as $M$ increases.
This shows that an LIS-based large antenna-array system is subject to the \emph{channel hardening effect} resulting in several practical implications.
First, an LIS-based system lacks scheduling diversity given that the fluctuations of the mutual information are small.
Further, an LIS offers an improved reliability insofar as it has a nearly deterministic data rate. 
Also, an LIS provides a low latency of having a deterministic data rate.
Furthermore, Theorem 2 shows that the ergodic rate of an LIS converges to the asymptotic bound ${\hat \mu _{{R_k}}}$ as $M$ increases.
We can observe that ${\hat \mu _{{R_k}}}$ is a function of ${\hat \mu _{{I_k}}}$ which depends exclusively on ${\left| {{\boldsymbol{h}}_{kk}^{\rm{H}}{\boldsymbol{h}}_{jk}^{\rm{L}}} \right|^2}$.
Therefore, the asymptotic bound is only affected by the interference signals through the LOS path from other devices. 
\emph{Hardware impairments, noise, and interference from estimation errors and the NLOS path become negligible} compared to LOS interference as $M$ increases. 
If all of the interference is generated from the NLOS path, the asymptotic bound goes to infinity as $M$ increases. 
Moreover, Lemma 5 and Theorem 2 show that the approximation gap resulting from the Taylor expansions in Corollary 1 goes to zero as $M\to\infty$.
As $M$ increases, ${\bar \mu _{{I_k}}}$ and $\bar \sigma _{{I_k}}^2$ follow, respectively, $\mathcal{O}(M^2)$ and $\mathcal{O}(M^3)$ as proved in Lemma 5,
and thus, $\bar \sigma _{{I_k}}^2/{\bar \mu _{{I_k}}^2}$ follows $\mathcal{O}(1/M)$ and eventually converges to zero.
Hence, since the terms of a higher order than the second degree of the Taylor expansion become negligible compared to the first and second-order terms,
the approximation gap resulting from Taylor expansions in (\ref{eq.MSINRk}) and (\ref{eq.SSINRk}) goes to zero as $M\to\infty$.
\begin{figure}[!ht]
\centering
\includegraphics[width=0.6\columnwidth] {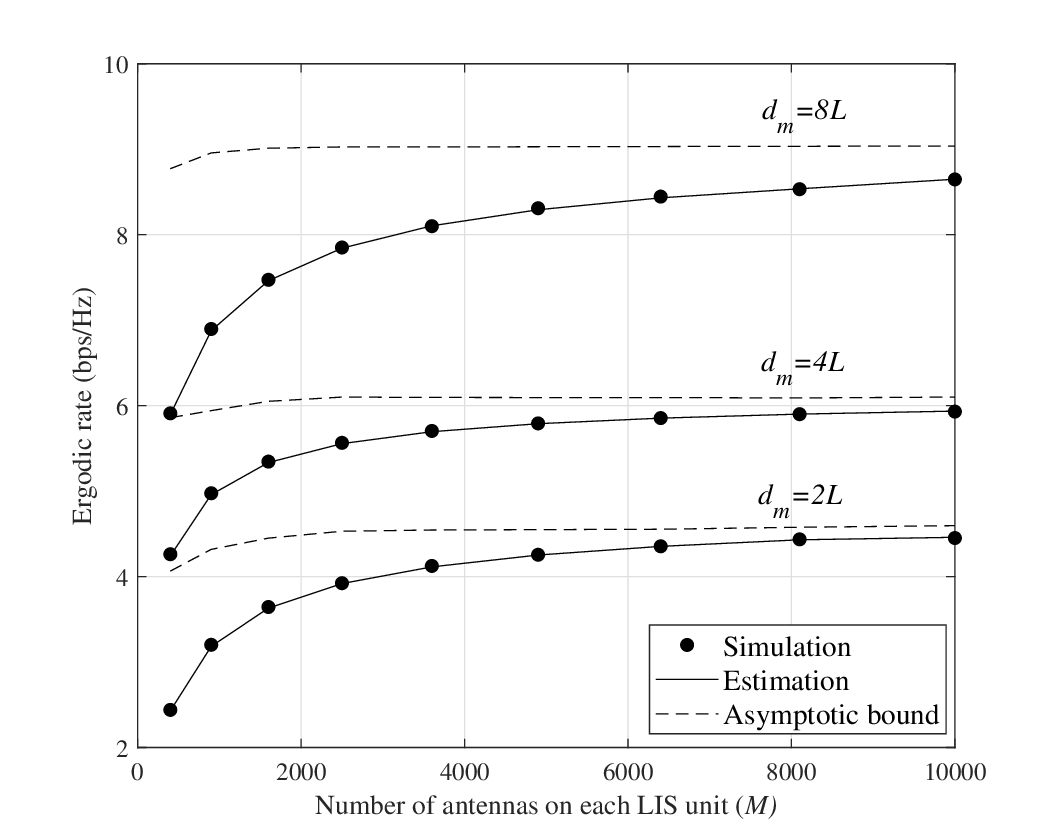}
\caption{{Ergodic rates of an LIS-based system with LOS interference as a function of the number of antennas on the LIS unit. }}
\label{fig.2}
\end{figure}
Similarly, $\bar\sigma _{{\gamma _k}}^2$ goes to zero as proved in Theorem 2 and the gap in (\ref{eq.MRk0}) and (\ref{eq.SRk0})
eventually goes to zero, as $M\to\infty$.

\section{Simulation Results and Analyses}
In this section, simulation results for the uplink rate in an LIS-based large antenna-array system are presented under a practical-sized environment with finite $M$ and $K$. 
Further, the asymptotic analyses are compared with the numerical results obtained from Monte Carlo simulations
(all simulations are statistically averaged over a large number of independent runs).
The simulation parameters are provided in Table \ref{table.para} and 
we do not consider shadowing given that the desired channel of LIS can be modeled as a perfect LOS.
In our results, the labels ``Estimation'' and ``Asymptotic bound'' refer to the results obtained from Theorems 1 and 2, respectively,
while the label ``Simulation'' captures a practical, simulated deployment of the considered LIS system.

\begin{figure}[!ht]
\centering
\includegraphics[width=0.6\columnwidth] {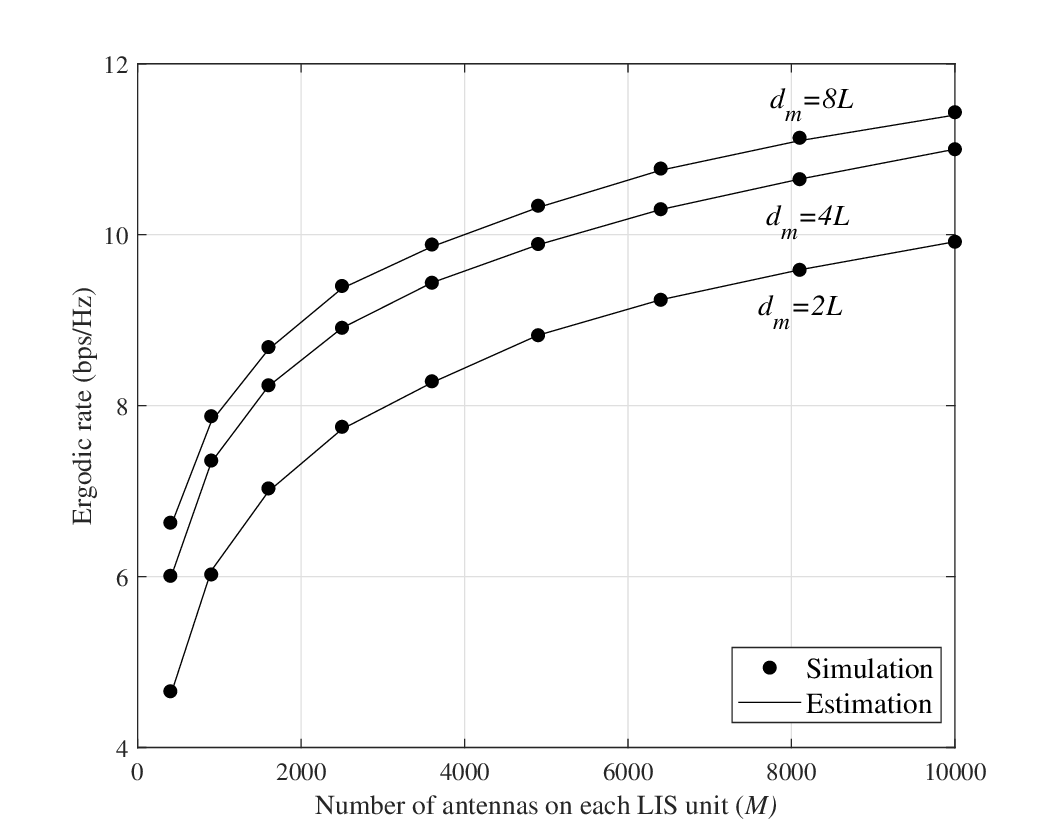}
\caption{{Ergodic rates of an LIS-based system with NLOS interference as a function of the number of antennas on the LIS unit.}}
\label{fig.3}
\end{figure}

In Figs. \ref{fig.2} and \ref{fig.3}, Theorems 1 and 2 are verified in the following scenario.
The devices are located at $z=1$ in parallel with the LIS on a two-dimensional plane. 
The devices are located in the ranges of $-10$ $\le x \le$ $10$ and $-10$ $\le y \le$ $10$ (in meters). 
The distance between the adjacent devices is set equally to ${d_{\rm{m}}}$ and the target device is located at $\left( {0,0,1} \right)$. 
Therefore, a total of 1681, 441, and 121 devices are located in a two-dimensional rectangular lattice form 
when ${d_{\rm{m}}} = 2L$, ${d_{\rm{m}}} = 4L$, and ${d_{\rm{m}}} = 8L$, respectively.

Figs. \ref{fig.2} and \ref{fig.3} compare the ergodic rates resulting from the simulations to the estimations from Theorem 1 as $M$ increases. 
In Fig. \ref{fig.2}, we assume that every interference signal from the other devices is generated entirely from the LOS path
and in Fig. \ref{fig.3}, it is generated entirely from the NLOS path. 
As shown in Figs. \ref{fig.2} and \ref{fig.3}, the asymptotic mean values derived from Theorem 1 are close to the results of our simulations over the entire range of $M$. 
We can also observe from Fig. \ref{fig.2} that the ergodic rate converges to the asymptotic bound obtained from Theorem 2 as $M$ increases.
However, in Fig. \ref{fig.3}, the ergodic rate goes to infinity as $M$ increases without bound. 
As proved in Theorem 2, we can see that only the interference stemming from a LOS path affects the ergodic rate of an LIS-based system.

\begin{figure}[!ht]
\centering
\includegraphics[width=0.6\columnwidth] {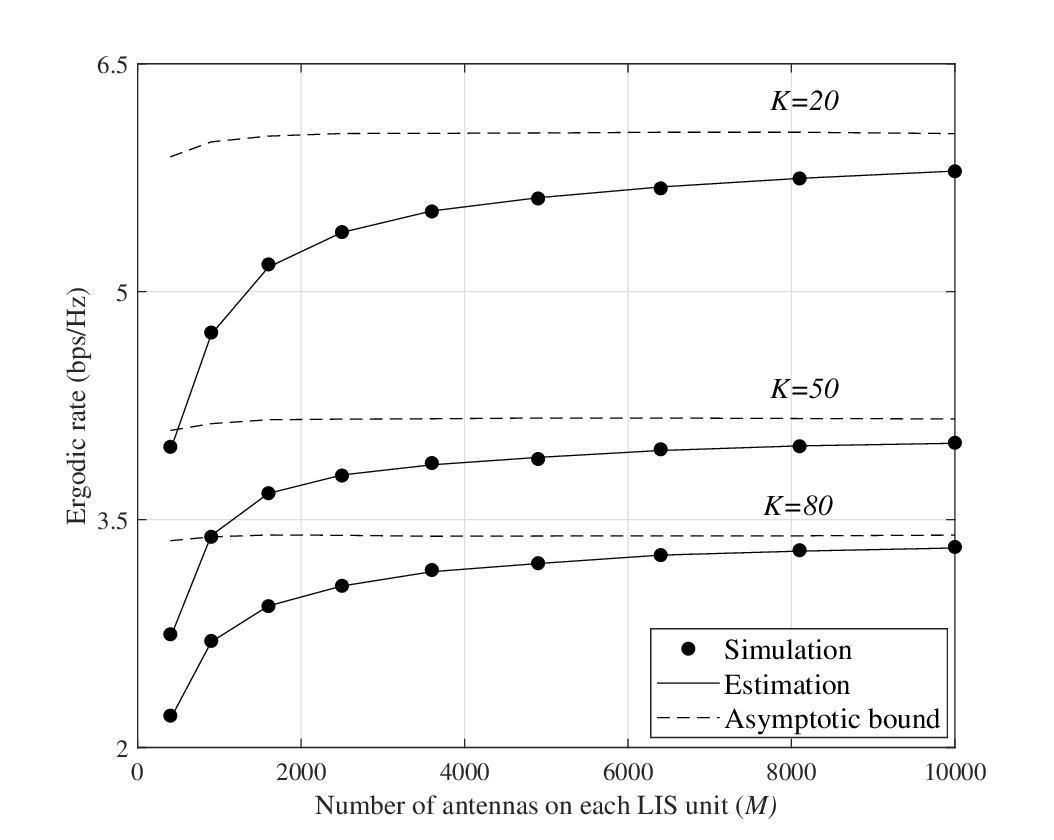}
\caption{{Ergodic rates of an LIS-based system with randomly located devices as a function of the number of antennas on the LIS unit.}}
\label{fig.4}
\end{figure}

\begin{figure}[!ht]
\centering
\includegraphics[width=0.6\columnwidth] {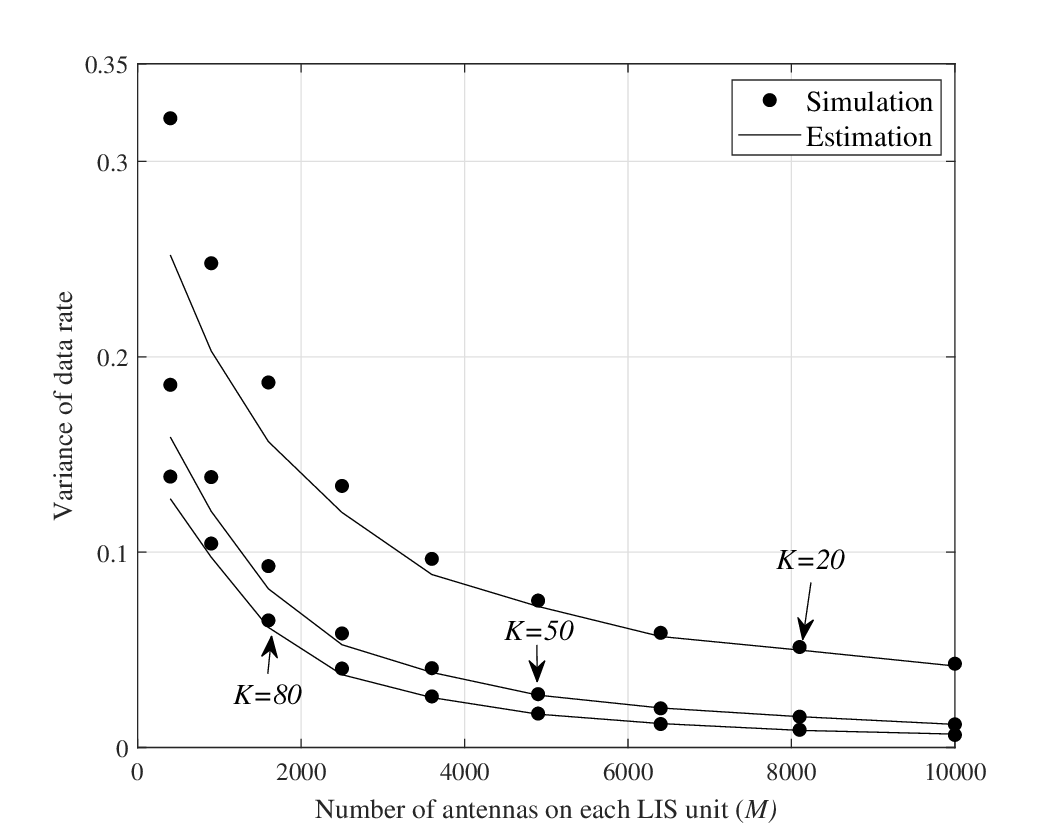}
\caption{{Variances of uplink rates of an LIS-based system with randomly located devices as a function of the number of antennas on the LIS unit.}}
\label{fig.5}
\end{figure}

\begin{figure}[!ht]
\centering
\includegraphics[width=0.6\columnwidth] {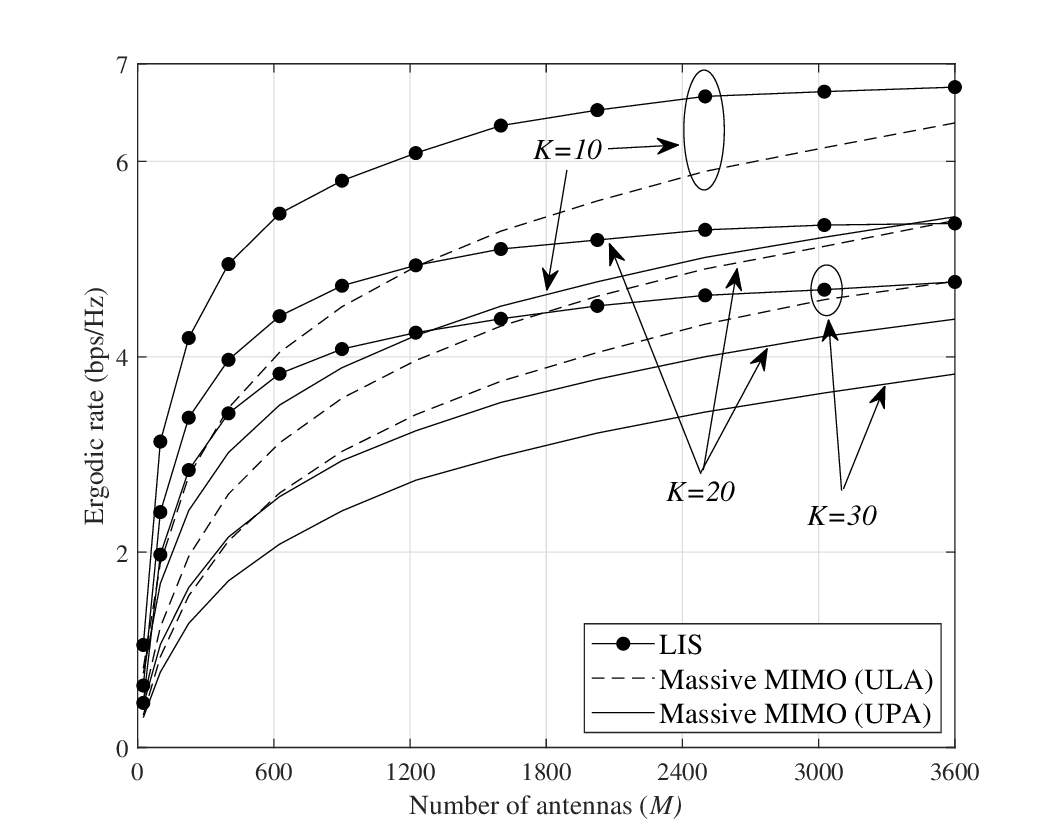}
\caption{{Performance comparison between the ergodic rates of an LIS-based system and a massive MIMO system as a function of the number of antennas. }}
\label{fig.6}
\end{figure}

\begin{figure}[!ht]
\centering
\includegraphics[width=0.6\columnwidth] {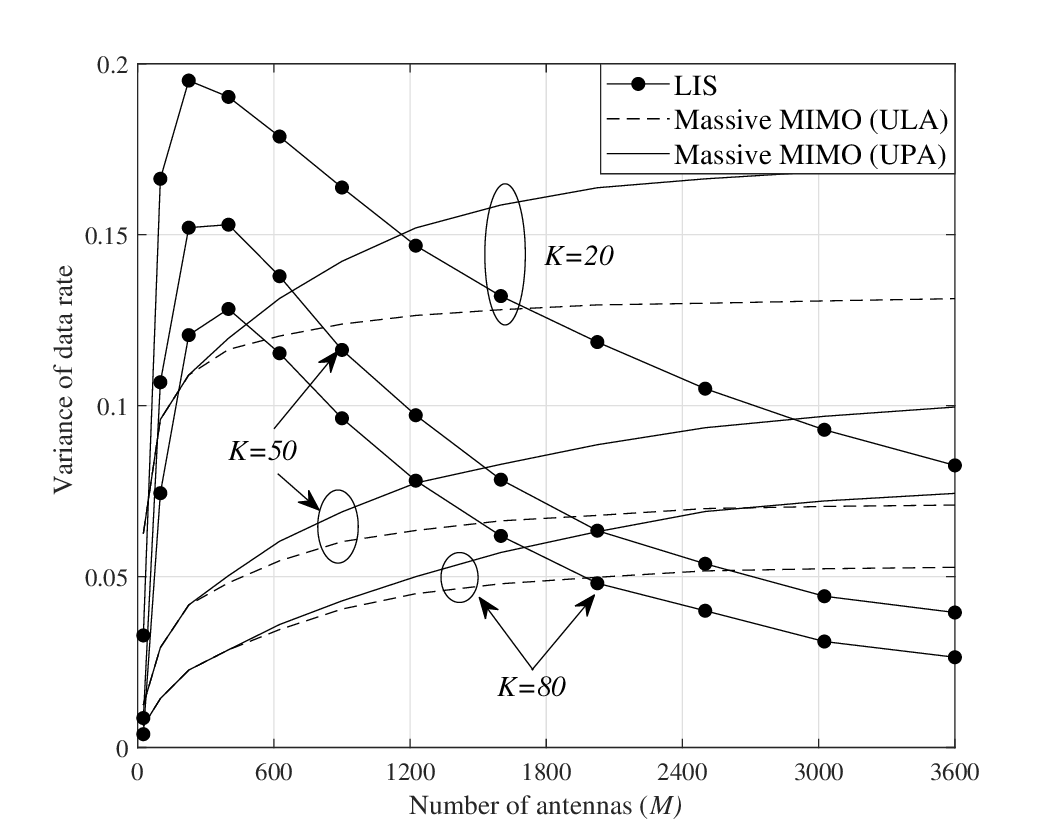}
\caption{{Performance comparison between variances of uplink rates resulting from an LIS-based system and a massive MIMO system as a function of the number of antennas. }}
\label{fig.7}
\end{figure}

\begin{figure}[!ht]
\centering
\includegraphics[width=0.6\columnwidth] {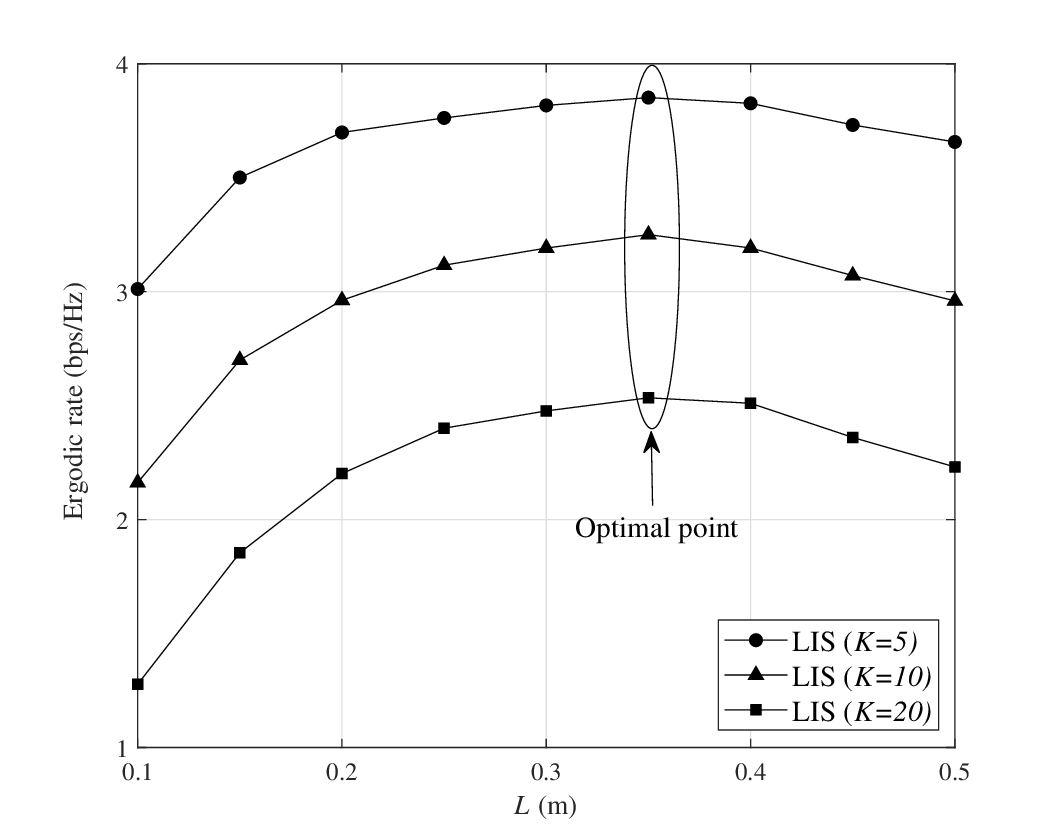}
\caption{{Ergodic rates of an LIS-based system as a function of $L$ when $M=100$. }}
\label{fig.8}
\end{figure}

In Figs. \ref{fig.4}--\ref{fig.8}, we consider that the devices are uniformly distributed within a three-dimensional space.
In particular, these figures are generated for a scenario in which we randomly and uniformly deploy the devices in a $4$ m$\times$ $4$ m $\times$ $2$ m space.
Based on the 3GPP model in \cite{ref.LTE}, the existence of a LOS path depends on the distance from the transmitter and receiver.
The probability of LOS is then given as follows:
\begin{equation}
P_{jk}^{{\rm{LOS}}} = \left\{ \begin{array}{l}
\left( {{d_{\rm{C}}} - {d_{jk}}} \right)/{d_{\rm{C}}},{\rm{ }}0 < {d_{jk}} < {d_{\rm{C}}},\\
\,{\qquad\quad}0,\quad\qquad{d_{jk}} > {d_C},
\end{array} \right.
\end{equation}
where ${d_{jk}}$ is the distance in meters between device $j$ and the center of LIS unit $k$, and ${d_{\rm{C}}}$ denotes a cutoff point, which is typically set to $300$ m in a cellular environment \cite{ref.LTE}. 
Since the antenna of a BS in a cellular system is located at a high altitude, ${d_{\rm{C}}}$ takes a large value such as $300$ m.
However, in an LIS environment, a relatively smaller ${d_{\rm{C}}}$ value is more reasonable. 
In Figs. \ref{fig.4}--\ref{fig.8}, ${d_{\rm{C}}} = 10$ m is assumed to jointly consider the LOS and NLOS path simultaneously. 
If a LOS path occurs, the Rician factor, $\kappa_{jk}$, is calculated according to ${d_{jk}}$ as per in Table \ref{table.para}.

Figs. \ref{fig.4} and \ref{fig.5} show the ergodic rate and variance of the rate, respectively, as a function of $M$. 
The asymptotic mean and variance derived from Theorem 1 are close to the results of our simulations 
and the accuracy improves as $M$ increases.
In Fig. \ref{fig.4}, the asymptotic means closely approximate the results of the simulations regardless of $K$,
whereas the asymptotic variances approach the results of the simulations as $K$ increases as shown in Fig. \ref{fig.5}.
Based on Theorem 1, the gap between the actual mean and asymptotic mean approaches zero as $M$ increase,
while the gap between the actual variance and asymptotic variance approaches zero as both $M$ and $K$ increase.
Fig. \ref{fig.4} also shows that the ergodic rate gradually converges to the asymptotic bound obtained from Theorem 2. 
Given that the interference power increases as $K$ increases, the ergodic rate gradually decreases as $K$ increases.

Fig. \ref{fig.5} shows the channel hardening effect whereby the rate variance gradually converges to zero as $M$ increases.
Moreover, the rate variance gradually decreases as $K$ increases for a fixed $M$.
From the scaling laws of ${\bar \mu _{{I_k}}}$ and $\bar \sigma _{{I_k}}^2$ in Lemma 4, 
${\bar \mu _{{I_k}}}$ and $\bar \sigma _{{I_k}}^2$ follow $\mathcal{O}\left( K \right)$ and $\mathcal{O}\left( {{K^2}} \right)$, respectively.
Then, $\bar \sigma _{{R_k}}^2$ follows $\mathcal{O}\left( {1/{K^2}} \right)$ from (\ref{eq.SRk0})--(\ref{eq.SSINRk}).
Therefore, $\bar \sigma _{{R_k}}^2$ decreases as $K$ increases. 

In Figs. \ref{fig.6} and \ref{fig.7}, we compare the performances of an LIS-based large antenna-array system and a massive MIMO system.
We consider a multi-user massive MIMO system in which an MF is used for uplink signal detection.
Since massive MIMO systems typically operate via far-field communications, we assume that every wireless signal is from an NLOS path and the distance from a device to all BS antennas is taken as equal \cite{ref.How}. 
For a massive MIMO system with a uniform linear array (ULA), the wireless channels from the NLOS path can be modeled using (\ref{eq.hjk}) with ${\kappa _{jk}} = 0$ and
\begin{equation}
{\boldsymbol{d}}\left( \theta _{jkp}^{\rm{h}} \right) = \frac{1}{{\sqrt M }}{\left[ {\begin{array}{*{20}{c}}
{1,}&{{e^{j\frac{{2\pi \nu }}{\lambda }\sin \theta _{jkp}^{\rm{h}}}},}& \cdots &{{e^{j\frac{{2\pi \nu}}{\lambda }\left( {M - 1} \right)\sin \theta _{jkp}^{\rm{h}}}}}
\end{array}} \right]^{\rm{T}}},
\end{equation}
which is then applied to (\ref{eq.DNLOS}). 
Here, $\nu$ is the antenna spacing of a massive MIMO system assuming $\nu=\lambda /2$.
We also assume a single BS with $M$ antennas and $P = M/2$, as in \cite{ref.How},
and the same device distribution is considered as in the case of the LIS.
For a fair comparison, we assume that the antenna gain is always equal to $1$ in both cases (i.e., with massive MIMO and the LIS).

Fig. \ref{fig.6} compares the ergodic rates of an LIS-based large antenna-array system and a massive MIMO system as $M$ increases.
This figure shows that the ergodic rates resulting from LIS are higher than those resulting from massive MIMO in the range of practical-sized $M$, since the desired signal power of the LIS channel (i.e., LOS channel) is higher than that of the massive MIMO channel (i.e., NLOS channel).
The performance gap decreases as $M$ increases because the interference signal from the NLOS path becomes negligible, and eventually the massive MIMO system becomes an interference-free environment.
When $K=30$, an LIS shows {about 2-fold} increase in the ergodic rate compared to massive MIMO with ULA at $M=100$,
but two systems achieve a nearly equal ergodic rate at $M=3600$.
However, the increase of $M$ indicates an increase in the physical area for deploying the massive antennas, 
whereas the physical area of the LIS remains constant at $2L \times 2L$. 
For example, the ergodic rates resulting from the LIS and the massive MIMO systems are almost equal when $K = 30$ and $M = 3600$.
The total physical length of the massive MIMO antennas is equal to $180$ m under the assumption of a ULA with $\lambda /2$-spacing. 
Even if we consider a two-dimensional antenna deployment, a $60 \times 60$ antenna-array occupies an area of roughly $9$ $\rm{m}^2$. 
However, the LIS unit only occupies an area of $0.25$ $\rm{m}^2$.
An interference-free MIMO environment is practically impossible given that the size of its array would have to be tremendous.
This clearly shows the advantages of an LIS for space-intensive wireless communication.

Fig. \ref{fig.7} compares the variances of uplink rates resulting from an LIS-based large antenna-array system and a massive MIMO system as $M$ increases.
In Fig. \ref{fig.7}, we plot the rate variance of LIS using the estimated value obtained from Theorem 1.
We can observe that the rate variance of massive MIMO increases as $M$ increases and then eventually converge to constant value exemplifying the so-called reduced channel hardening effect \cite{ref.Hard}.
However, the rate variance of LIS converges to zero as $M$ increases due to the channel hardening effect.
Therefore, an LIS has improved reliability having a deterministic rate and it results in lower latency compared to a massive MIMO system.

Fig. \ref{fig.8} shows the ergodic rates resulting from an LIS-based large antenna-array system as a function of $L$ when $M=100$.
In the LIS, the maximum SINR is achieved at the central antenna of the LIS unit and the SINR gradually decreases as the antenna moves from the center to the edge.
Thus, the ergodic rate increases as $L$ increases when $L$ is small and decreases when $L$ exceeds some threshold point.
As shown in Fig. \ref{fig.8}, the maximum ergodic rates can be achieved through optimal $L$ values.
Furthermore, an optimal $L$ can be obtained numerically using the asymptotic analysis from Theorem 1 as given by {$L=0.35$ m} in Fig. \ref{fig.8}.

\section{Conclusions}
In this paper, we have asymptotically analyzed the uplink data rate of an LIS-based large antenna-array system. 
We have derived the asymptotic moments of mutual information by considering a practical LIS environment in which a large LIS can be divided into smaller LIS units, each of which having a limited area.
We have studied the uplink rate in presence of limitations such as {hardware impairments,} imperfect channel estimation, and interference that is generated by device-specific spatially correlated Rician fading.
We have shown that our analyses can accurately determine the performance of an LIS analytically, without the need for extensive simulations.
Furthermore, we have demonstrated that a channel hardening effect will occur in an LIS-based system.
We have also derived the asymptotic bound of the uplink data rate and shown that {hardware impairments,} noise, and interference from channel estimation errors and the NLOS path become negligible as $M$ increases. 
The simulation results have shown that the results of our asymptotic analyses agree with those resulting from extensive simulations, and the ergodic rate and the variance of rate respectively converge to the derived asymptotic bound and zero as $M$ and $K$ increase.
Moreover, we have observed that an LIS enables reliable and space-intensive communication, which renders it a promising technology beyond massive MIMO systems.
We expect that our asymptotic analyses will be invaluable to predict the theoretical performance of an LIS-based large antenna-array system when conducting system-level simulations and developing prototypes.

\section*{Appendix A\\ Proof of Lemma 1}
Given the definition of ${X_k} = {\left| {{\boldsymbol{e}}_k^{\boldsymbol{H}}{{\boldsymbol{h}}_{kk}}} \right|^2}$ from (\ref{eq.Xk}), we have
\begin{equation}
{X_k} = {\left| {\sum\limits_m^M {{{\left( {\beta _{km}^{\rm{L}}} \right)}^2}e_{km}^*{h_{kkm}}} } \right|^2}.
\end{equation}
Let us define ${\widetilde X_{km}} = {\left( {\beta _{km}^{\rm{L}}} \right)^2}e_{km}^*{h_{kkm}}$ $\forall k,m$.
Then, ${\widetilde X_{km}} \sim \mathcal{CN}\left( {0,{{\left( {\beta _{km}^{\rm{L}}} \right)}^4}} \right)$ 
and ${X_k}$ can be described as follows: 
\begin{equation}
{X_k} \sim \frac{1}{2}{{\sum\limits_m^M {{{\left( {\beta _{km}^{\rm{L}}} \right)}^4}} }}\chi _2^2, 
\end{equation}
where $\chi _k^2$ denotes the chi-square distribution with $k$ degrees of freedom, which completes the proof. 

\section*{Appendix B\\ Proof of Lemma 2}
Given the definition of ${Y_{jk}} = {\left| {\sqrt {1 - \tau _k^2} {\boldsymbol{h}}_{kk}^{\rm{H}}{{\boldsymbol{h}}_{jk}} + {\tau _k}{\boldsymbol{e}}_k^{\rm{H}}{{\boldsymbol{h}}_{jk}}} \right|^2}$ from (\ref{eq.Yjk}), we have
\begin{equation}
{Y_{jk}} = {\left| {{\Sigma _{{Y_{jk}}}}} \right|^2} = {\left| {Y_{jk}^{\rm{L}} + Y_{jk}^{{\rm{N1}}} + Y_{jk}^{{\rm{N2}}}} \right|^2}, \label{eq.B.Y}
\end{equation}
where we define ${{{\Sigma _{{Y_{jk}}}}} } = {Y_{jk}^{\rm{L}} + Y_{jk}^{{\rm{N1}}} + Y_{jk}^{{\rm{N2}}}}$ $\forall j,k$ and
\begin{align}
Y_{jk}^{\rm{L}} &= \sqrt {\frac{{{\kappa _{jk}}}}{{{\kappa _{jk}} + 1}}} \left( {\sqrt {1 - \tau _k^2} {\boldsymbol{h}}_{kk}^{\rm{H}}{\boldsymbol{h}}_{jk}^{\rm{L}}} 
+ {{\tau _k}\sum\limits_m^M {\beta _{km}^{\rm{L}}\beta _{jm}^{\rm{L}}e_{km}^*{h_{jkm}}} } \right), \forall j,k \\
Y_{jk}^{{\rm{N1}}} &= \sqrt {\frac{1}{{{\kappa _{jk}} + 1}}} \left( {\sqrt {1 - \tau _k^2} {\boldsymbol{h}}_{kk}^{\rm{H}}{\boldsymbol{R}}_{jk}^{1/2}{{\boldsymbol{g}}_{jk}}} \right) 
= \sqrt {\frac{{1 - \tau _k^2}}{{{\kappa _{jk}} + 1}}} \sum\limits_p^P {{\boldsymbol{h}}_{kk}^{\rm{H}}{{\boldsymbol{r}}_{jkp}}{g_{jkp}}} , \forall j,k\\
Y_{jk}^{{\rm{N2}}} &= \sqrt {\frac{1}{{{\kappa _{jk}} + 1}}} \left( {{\tau _k}{\boldsymbol{e}}_k^{\rm{H}}{\boldsymbol{R}}_{jk}^{1/2}{{\boldsymbol{g}}_{jk}}} \right) 
= \sqrt {\frac{{\tau _k^2}}{{{\kappa _{jk}} + 1}}} \sum\limits_{m,p}^{M,P} {\beta _{km}^{\rm{L}}{r_{jkmp}}e_{km}^*{g_{jkp}}}, \forall j,k \label{eq.B.YN2}
\end{align}
where $\boldsymbol{R}_{jk}^{1/2} = \left[ {{{\boldsymbol{r}}_{jk1}}, \ldots ,{{\boldsymbol{r}}_{jkP}}} \right]$ and ${{\boldsymbol{r}}_{jkp}} = {\left[ {{r_{jk1p}}, \ldots ,{r_{jkMp}}} \right]^{\rm{T}}} $ $\forall j,k,p$.
We first prove that $Y_{jk}^{\rm{L}}$ and $Y_{jk}^{\rm{N1}}$ follow independent complex Gaussian distributions.
We then asymptotically obtain the distribution of $Y_{jk}^{\rm{N2}}$ by using the law of large numbers for a large $M$ and the Lyapunov central limit theorem (CLT).
Finally, we asymptotically derive the mean and variance of $Y_{jk}$.
Given that $e_{km}$, $g_{jkp}$, and $e_{km}g_{jkp}$ are independent of each other, $Y_{jk}^{{\rm{L}}}$, $Y_{jk}^{{\rm{N1}}}$, and $Y_{jk}^{{\rm{N2}}}$ are independent random variables.
Further, since $e_{km}$ and $g_{jkp}$ are standard complex Gaussian random variables respectively independent across $m$ and $p$,
we have
$Y_{jk}^{\rm{L}} \sim \mathcal{CN}\left( {\mu _{jk}^{\rm{L}}},s_{jk}^{\rm{L}} \right)$ and
$Y_{jk}^{{\rm{N1}}}\sim \mathcal{CN}\left( {0,s_{jk}^{\rm{N1}}} \right)$, 
where
\begin{align}
{\mu _{jk}^{\rm{L}}} &= \sqrt {\frac{{{\kappa _{jk}}\left( {1 - \tau _k^2} \right)}}{{{\kappa _{jk}} + 1}}} {\boldsymbol{h}}_{kk}^{\rm{H}}{\boldsymbol{h}}_{jk}^{\rm{L}}, \label{eq.B.ML} \\
s_{jk}^{\rm{L}} &= \frac{{{\kappa _{jk}}\tau _k^2}}{{{\kappa _{jk}} + 1}}\sum\limits_m^M {{{\left( {\beta _{km}^{\rm{L}}\beta _{jm}^{\rm{L}}} \right)}^2}} , \label{eq.B.SL} \\
s_{jk}^{\rm{N1}} &= \frac{{1 - \tau _k^2}}{{{\kappa _{jk}} + 1}}{\sum\limits_p^P {\left| {{\boldsymbol{h}}_{kk}^{\rm{H}}{{\boldsymbol{r}}_{jkp}}} \right|} ^2}. \label{eq.B.SN1}
\end{align}
Here, ${\boldsymbol{h}}_{kk}^{\rm{H}}{\boldsymbol{h}}_{jk}^{\rm{L}}$ in (\ref{eq.B.ML}), ${\sum\nolimits_m^M {\left( {\beta _{km}^{\rm{L}}\beta _{jm}^{\rm{L}}} \right)} ^2}$ in (\ref{eq.B.SL}), and ${\sum\nolimits_p^P {\left| {{\boldsymbol{h}}_{kk}^{\rm{H}}{{\boldsymbol{r}}_{jkp}}} \right|} ^2}$ in (\ref{eq.B.SN1}) are deterministic values depending on the locations of the devices and the correlation matrices.

In order to obtain a random variable $Y_{jk}^{{\rm{N2}}}$, we use the law of large numbers to approximate (\ref{eq.B.YN2}) for a large $M$. 
$\sum\nolimits_m^M {\beta _{km}^{\rm{L}}{r_{jkmp}}e_{km}^*{g_{jkp}}} $ in (\ref{eq.B.YN2}) is thus expressed as follows: 
\begin{equation}
\sum\limits_m^M {\beta _{km}^{\rm{L}}{r_{jkmp}}e_{km}^*{g_{jkp}}}  = \alpha _{jkp}^{{\rm{NL}}}\sum\limits_m^M {\beta _{km}^{\rm{L}}l_{jkm}^{{\rm{NL}}}{d_{jkmp}}e_{km}^*{g_{jkp}}}, 
\end{equation}
where ${d_{jkmp}}$ is element $m$ of ${\boldsymbol{d}}\left( {\phi _{jkp}^{\rm{v}},\phi _{jkp}^{\rm{h}}} \right)$.
We define $\widetilde{Y}_{jkp} = \sum\nolimits_m^M {\beta _{km}^{\rm{L}}l_{jkm}^{{\rm{NL}}}{d_{jkmp}}{y_{jkmp}}}$ $\forall j,k,p$ where ${y_{jkmp}} = e_{km}^*{g_{jkp}}$.
Then, the random variable $\widetilde{Y}_{jkp}$ follows Corollary 2.

{\bf{{Corollary 2.}}} On the basis of the Lyapunov CLT, a random variable $\widetilde{Y}_{jkp}$ asymptotically follows a complex Gaussian distribution:
\begin{equation}
\sqrt{\frac{M}{\sum\nolimits_m^M {{{\left( {\beta _{km}^{\rm{L}}l_{jkm}^{{\rm{NL}}}} \right)}^2}}}}\widetilde{Y}_{jkp}\xrightarrow[M \to \infty ]{\rm{d}} \mathcal{CN}\left( {0,1 } \right),
\end{equation}
where `` $\xrightarrow[M \to \infty ]{\rm{d}}$'' denotes the convergence in distribution.

\begin{proof}
In order to follow the Lyapunov CLT, ${y_{jkmp}}$ should be independent random variable across $m$ and the following Lyapunov's condition should be satisfied for some $\delta  > 0$ \cite{ref.CLT}:
\begin{equation}
\mathop {\lim }\limits_{M \to \infty } \frac{1}{{s_M^{2 + \delta }}}\sum\limits_m^M {{\rm{E}}\left[ {{{\left| {\beta _{km}^{\rm{L}}l_{jkm}^{{\rm{NL}}}{d_{jkmp}}{y_{jkmp}} - {\mu _m}} \right|}^{2 + \delta }}} \right]}  = 0, \label{eq.B.CLTcondition}
\end{equation}
where $s_M^2=\sum\nolimits_m^M {\sigma _m^2}$, and ${\mu _m}$ and $\sigma _m^2$ are the mean and variance of the random variable $\beta _{km}^{\rm{L}}l_{jkm}^{{\rm{NL}}}{d_{jkmp}}{y_{jkmp}}$, respectively.
Here, ${y_{jkmp}} = e_{km}^*{g_{jkp}}$ is a random variable product of two independent random variables.
Since $e_{km}$ is an independent random variable across $m$ and independent with ${g_{jkp}}$, 
$\left\{ {{y_{jk1p}},...,{y_{jkMp}}} \right\}$ is a sequence of independent random variables, each with zero mean and unit variance.
Then, we can obtain ${\mu _m} = 0$ and $\sigma _m^2 = {\left( {\beta _{km}^{\rm{L}}l_{jkm}^{{\rm{NL}}}} \right)^2}{\left| {{d_{jkmp}}} \right|^2} = {\left( {\beta _{km}^{\rm{L}}l_{jkm}^{{\rm{NL}}}} \right)^2}/M$ $\forall m$.
We consider $\delta = 2$, such that (\ref{eq.B.CLTcondition}) is obtained as follows:
\begin{equation}
\mathop {\lim }\limits_{M \to \infty } \frac{{\sum\limits_m^M {{{\left( {\beta _{km}^{\rm{L}}l_{jkm}^{{\rm{NL}}}} \right)}^4}} {\rm{E}}\left[ {{{\left| {{y_{jkmp}}} \right|}^4}} \right]}}{{{{\left( {\sum\limits_m^M {{{\left( {\beta _{km}^{\rm{L}}l_{jkm}^{{\rm{NL}}}} \right)}^2}} } \right)}^2}}}\mathop  = \limits_{\left( a \right)} \mathop {\lim }\limits_{M \to \infty } \frac{{4\sum\limits_m^M {{{\left( {\beta _{km}^{\rm{L}}l_{jkm}^{{\rm{NL}}}} \right)}^4}} }}{{{{\left( {\sum\limits_m^M {{{\left( {\beta _{km}^{\rm{L}}l_{jkm}^{{\rm{NL}}}} \right)}^2}} } \right)}^2}}}
= \mathop {\lim }\limits_{M \to \infty } \frac{{4\bar \alpha_{jk} }}{{M{{\tilde \alpha_{jk} }^2}}},\label{eq.B.CLTcondition2}
\end{equation}
where we define $\bar \alpha_{jk}  = \sum\nolimits_m^M {{{\left( {\beta _{km}^{\rm{L}}l_{jkm}^{{\rm{NL}}}} \right)}^4}} /M$ and $\tilde \alpha_{jk}  = \sum\nolimits_m^M {{{\left( {\beta _{km}^{\rm{L}}l_{jkm}^{{\rm{NL}}}} \right)}^2}} /M$ $\forall j,k$.
(a) results from ${\rm{E}}\left[ {{{\left| {{y_{jkmp}}} \right|}^4}} \right]=4$ $\forall m$ since ${y_{jkmp}}$ is product of two independent random variables that follow an identical standard complex Gaussian distribution.
Given that $0 < \beta _{km}^{\rm{L}} \le 1$ and $0 < l_{jkm}^{{\rm{NL}}}  \le 1$, we have $0 < {\bar \alpha_{jk} }\le 1$ and $0 < {\tilde \alpha_{jk} }\le 1$.
Therefore, (\ref{eq.B.CLTcondition2}) goes to zero, which completes the proof.
\end{proof}

Based on Corollary 2, we have $\frac{1}{\sqrt{s_{jk}^{{\rm{N2}}}}}Y_{jk}^{{\rm{N2}}}\xrightarrow[M \to \infty ]{\rm{d}}\mathcal{CN}\left( {0,1} \right)$, where 
\begin{equation}
s_{jk}^{{\rm{N2}}} = \frac{{\tau _k^2}}{{{\kappa _{jk}} + 1}}\sum\limits_{m,p}^{M,P} {{{\left( {\alpha _{jkp}^{{\rm{NL}}}\beta _{km}^{\rm{L}}l_{jkm}^{{\rm{NL}}}} \right)}^2}} /M. \label{eq.B.SN2}
\end{equation}
Here, $\alpha _{jkp}^{{\rm{NL}}}\beta _{km}^{\rm{L}}l_{jkm}^{{\rm{NL}}}$ is a deterministic value depending on the locations of the devices and the correlation matrices.
Given that $Y_{jk}^{{\rm{L}}}$, $Y_{jk}^{{\rm{N1}}}$, and $Y_{jk}^{{\rm{N2}}}$ are independent of each other,
we have $\frac{1}{\sqrt{s_{jk}^{\rm{L}} + s_{jk}^{\rm{N1}} + s_{jk}^{{\rm{N2}}}}}\left({\Sigma _{{Y_{jk}}}}-{\mu _{jk}^{\rm{L}}}\right)\xrightarrow[M \to \infty ]{\rm{d}}\mathcal{CN}\left( 0,1 \right)$.
For a random variable $Y=\sum\nolimits_i {\left| {X_i} \right|}^2$, where $X_i$ is a complex Gaussian random variable independent across $i$ such as $X_i \sim \mathcal{CN}\left(m_i,\sigma _{i}^2 \right)$,
the mean and variance of $Y$ are respectively obtained by
$\mu_{Y} = \sum\nolimits_i \left({\sigma _{i}^2} + {\left| {{m _{i}}} \right|^2}\right)$ and
$\sigma_{Y}^2 = \sum\nolimits_i \left({\sigma _{i}^4} + 2{\left| {{m _{i}}} \right|^2}\sigma _{i}^2\right).$
Then, ${\bar \mu _{{Y_{jk}}}}$ and $\bar \sigma _{{Y_{jk}}}^2$ are ultimately obtained by
\begin{align}
{\bar \mu _{{Y_{jk}}}} &= s_{jk}^{\rm{L}} + s_{jk}^{\rm{N1}} + s_{jk}^{{\rm{N2}}} + {\left| {{\mu _{jk}^{\rm{L}}}} \right|^2}, \label{eq.B.MY}\\
\bar \sigma _{{Y_{jk}}}^2 &= {\left( {s_{jk}^{\rm{L}} + s_{jk}^{\rm{N1}} + s_{jk}^{{\rm{N2}}}} \right)^2} + 2{\left| {{\mu _{jk}^{\rm{L}}}} \right|^2}\left( {s_{jk}^{\rm{L}} + s_{jk}^{\rm{N1}} + s_{jk}^{{\rm{N2}}}} \right). \label{eq.B.SY}
\end{align}

\section*{Appendix C\\ Proof of Lemma 3}
Given the definition of ${Z_{k}}$ from (\ref{eq.Zk}), we have
\begin{align}
{{Z_k^{\rm{n}}}} &= {\left| {\sqrt {1 -    \tau _k^2} {\boldsymbol{h}}_{kk}^{\rm{H}} + {\tau _k}{\boldsymbol{e}}_k^{\rm{H}}} \right|^2} = \sum\limits_m^M {{{\left| {{{z_{km}^{\rm{n}}}}} \right|}^2}}, \\
{{Z_k^{\rm{w}}}} &{={{\left| {\sqrt {1 - \tau _k^2} {\boldsymbol{h}}_{kk}^{\rm{H}}{{\boldsymbol{\widetilde w}}_k} + {\tau _k}{\boldsymbol{e}}_k^{\rm{H}}}{{\boldsymbol{\widetilde w}}_k} \right|}^2}
={{\left| \sum\limits_m^M {z_{km}^{\rm{w}}} \right|}^2},}
\end{align}
where we define ${{z_{km}^{\rm{n}}}} = \sqrt {1 - \tau _k^2} \beta _{km}^{\rm{L}}h_{kkm}^* + {\tau _k}\beta _{km}^{\rm{L}}e_{km}^*$ {and ${z_{km}^{\rm{w}}} = \sqrt {1 - \tau _k^2} \beta _{km}^{\rm{L}}h_{kkm}^*{\widetilde w}_m + {\tau _k}\beta _{km}^{\rm{L}}e_{km}^*{\widetilde w}_m = {\widetilde w}_m{z_{km}^{\rm{n}}}$.} 
Then, ${{z_{km}^{\rm{n}}}}$ follows a complex Gaussian distribution:
\begin{equation}
{{z_{km}^{\rm{n}}}} \sim \mathcal{CN}\left( {\sqrt {1 - \tau _k^2} \beta _{km}^{\rm{L}}h_{kkm}^*,\tau _k^2{{\left( {\beta _{km}^{\rm{L}}} \right)}^2}} \right).
\end{equation}
Then, the mean and variance of ${Z_k^{\rm{n}}}$ are ${{\mu _{{Z_k^{\rm{n}}}}}} = \sum\limits_m^M {{{\left( {\beta _{km}^{\rm{L}}} \right)}^2}}$ and ${\sigma _{{Z_k^{\rm{n}}}}^2} = \tau _k^2\left( {2 - \tau _k^2} \right)\sum\limits_m^M {{{\left( {\beta _{km}^{\rm{L}}} \right)}^4}}$, respectively.
{
Similarly, as proved in Corollary 2, $\sum\nolimits_m^M {z_{km}^{\rm{w}}}$ follows a complex Gaussian distribution based on the Lyapunov CLT as $M \to \infty$.
From (\ref{eq.HWI}), ${z_{km}^{\rm{w}}}$ can be represented as follows:
\begin{equation}
{z_{km}^{\rm{w}}} = c_m{z_{km}^{\rm{n}}}\left(\sqrt{\rho_k}\beta _{km}^{\rm{L}}h_{kkm} + 
h_{km}^{\rm{R}}\right),\label{eq.C.zkm}
\end{equation}
where $h_{km}^{\rm{R}}=\sum\nolimits_{j \ne k}^K \frac{\sqrt{\rho_j} \left({\sqrt {{{{\kappa _{jk}}}}} {{\beta _{jm}^{\rm{L}}{h_{jkm}}} }+ {h_{jkm}^{\rm{NL}}}}\right)}{\sqrt{{\kappa _{jk}} + 1}}$.
We define ${z_{km}^{\rm{wL}}}=\sqrt{\rho_k}\beta _{km}^{\rm{L}}h_{kkm}c_m{z_{km}^{\rm{n}}}$
and ${z_{km}^{\rm{wR}}} = c_m{z_{km}^{\rm{n}}} h_{km}^{\rm{R}}$.
Since $c_m$ is a zero-mean complex Gaussian random variable inpendent with ${z_{km}^{\rm{n}}}$,
the mean and variance of ${z_{km}^{\rm{wL}}}$ are obtained from \cite{ref.Leo1960var}, respectively, as
$\mu_{{z_{km}^{\rm{wL}}}}=0$ and
\begin{equation}
\sigma_{{z_{km}^{\rm{wL}}}}^2= {\rho_k}(\beta _{km}^{\rm{L}})^2{\rm{Var}}\left[ c_m \right] \left({\big|{{\rm{E}}\left[ {z_{km}^{\rm{n}}} \right]}\big|^2 + {\rm{Var}}\left[ {z_{km}^{\rm{n}}} \right]}\right) = \delta\rho_k  (\beta _{km}^{\rm{L}})^4.\label{eq.C.SwL}
\end{equation}
Similary, the mean and variance of ${z_{km}^{\rm{wR}}}$ can be obtained, respectively, as $\mu_{{z_{km}^{\rm{wR}}}}=0$ and
\begin{equation}
\sigma_{{z_{km}^{\rm{wR}}}}^2  =  \delta \left({\Big|{\rm{E}}\left[{z_{km}^{\rm{n}}} h_{km}^{\rm{R}}\right]\Big|^2 }+ {{\rm{Var}}\left[{z_{km}^{\rm{n}}} h_{km}^{\rm{R}}\right] }\right).\label{eq.C.SwR}
\end{equation}
Since ${z_{km}^{\rm{n}}}$ and $h_{km}^{\rm{R}}$ are independent, we have
\begin{equation}
{\Big|{\rm{E}}\left[{z_{km}^{\rm{n}}} h_{km}^{\rm{R}}\right]\Big|^2 } =  ({1 - \tau _k^2})
\left|\sum\limits_{j \ne k}^K \sqrt{\frac{\rho_j {\kappa _{jk}}}{{{\kappa _{jk}} + 1}}}{\beta _{km}^{\rm{L}}{\beta _{jm}^{\rm{L}} h_{kkm}^* {h_{jkm}}} }\right|^2.\label{eq.C.E}
\end{equation}
Also, ${\rm{Var}}\left[{z_{km}^{\rm{n}}} h_{km}^{\rm{R}}\right]$ in (\ref{eq.C.SwR}) is obtained in a manner similar to (\ref{eq.C.SwL}), as follows:
\begin{equation}
{\rm{Var}}\left[{z_{km}^{\rm{n}}} h_{km}^{\rm{R}}\right] = 
(\beta _{km}^{\rm{L}})^2 \left( \sigma_{h_{km}^{\rm{R}}}^2 + \tau _k^2\big| \mu_{h_{km}^{\rm{R}}}\big|^2\right),\label{eq.C.Vzh}
\end{equation}
where
\begin{align}
\mu_{h_{km}^{\rm{R}}} &= \sum\limits_{j \ne k}^K \sqrt{\frac{\rho_j {\kappa _{jk}}}{{{\kappa _{jk}} + 1}}}{{\beta _{jm}^{\rm{L}} {h_{jkm}}} },\\
\sigma_{h_{km}^{\rm{R}}}^2 &= \sum\limits_{j \ne k}^K {\frac{\rho_j }{{{\kappa _{jk}} + 1}}}\sum\limits_p^P \left|{r_{jkmp}}\right|^2.
\end{align}
(\ref{eq.C.E}) and (\ref{eq.C.Vzh}) respectively show that ${\big|{\rm{E}}\left[{z_{km}^{\rm{n}}} h_{km}^{\rm{R}}\right]\big|^2 } $ and ${\rm{Var}}\left[{z_{km}^{\rm{n}}} h_{km}^{\rm{R}}\right]$ can be calculated based on the deterministic values such as the locations of the devices and the correlation matrices.
Since ${z_{km}^{\rm{wL}}}$ and ${z_{km}^{\rm{wR}}}$ include a common random variable $c_m z_{km}^{\rm{n}}$,
they are dependent of each other with the following covariance:
\begin{equation}
 \omega_{km}^{\rm{wLR}} =
{\rm{E}}\left[{{z_{km}^{\rm{wL}}}} \left({{z_{km}^{\rm{wR}}}}\right)^*\right]
=\delta \sqrt{\rho_k} {{\left( {\beta _{km}^{\rm{L}}} \right)}^3}
\sum\limits_{j \ne k}^K \sqrt{\frac{\rho_j {\kappa _{jk}}}{{{\kappa _{jk}} + 1}}}{{\beta _{jm}^{\rm{L}} h_{kkm} {h_{jkm}^*}} }.\label{eq.omega_km}
\end{equation}
Note that ${z_{km}^{\rm{wL}}}$ and ${z_{km}^{\rm{wR}}}$ are independent random variables across $m$.
Therefore, we can finally obtain the mean and variance of $Z_k^{\rm{w}}$ respectively as follows:
\begin{align}
\mu_{Z_k^{\rm{w}}}&=\sum\limits_m^M \left(\sigma_{{z_{km}^{\rm{wL}}}}^2 + \sigma_{{z_{km}^{\rm{wR}}}}^2 +
2 {\rm{Re}}\big( \omega_{km}^{\rm{wLR}} \big)\right),\label{eq.C.Mzw}\\
\sigma_{Z_k^{\rm{w}}}^2&= \left(\sum\limits_m^M \left(\sigma_{{z_{km}^{\rm{wL}}}}^2 + \sigma_{{z_{km}^{\rm{wR}}}}^2 + 2 {\rm{Re}}\big( \omega_{km}^{\rm{wLR}} \big)\right)\right)^2,\label{eq.C.Vzw}
\end{align}
which completes the proof.}

\section*{Appendix D\\ Proof of Lemma 4}
Given the definition of ${I_k} = {\rho _k}\tau _k^2{X_k} + \sum\nolimits_{j \ne k}^K {{\rho _j}{Y_{jk}}}  + {Z_k}$ from (\ref{eq.Ik_def}), we have
\begin{equation}
{I_k} = {\rho _k}\tau _k^2{\left| {{\boldsymbol{e}}_k^{\rm{H}}{{\boldsymbol{h}}_{kk}}} \right|^2} + \sum\nolimits_{j \ne k}^K {{\rho _j}{{\left| {\sqrt {1 - \tau _k^2} {\boldsymbol{h}}_{kk}^{\rm{H}}{{\boldsymbol{h}}_{jk}} + {\tau _k}{\boldsymbol{e}}_k^{\rm{H}}{{\boldsymbol{h}}_{jk}}} \right|}^2}} + {\left| {\sqrt {1 - \tau _k^2} {\boldsymbol{h}}_{kk}^{\rm{H}} + {\tau _k}{\boldsymbol{e}}_k^{\rm{H}}} \right|^2}.\label{eq.D.Ik}
\end{equation}
Based on Lemmas 1--3, ${\bar \mu _{{I_k}}}$ can be obtained as
${\bar \mu _{{I_k}}} = {\rho _k}\tau _k^2{\bar q_k} + \sum\nolimits_{j \ne k}^K {{\rho _j}{{\bar \mu }_{{Y_{jk}}}}}  + \sqrt {{{\bar p}_k}}$. 
We can observe from (\ref{eq.D.Ik}) that ${X_k}$, $\sum\nolimits_j {{\rho _j}{Y_{jk}}}$, and $Z_k$ are function of a common random variable vector ${{\boldsymbol{e}}_k}$. 
Similarly, ${Y_{ik}}$ and ${Y_{jk}}$ $\forall i\ne j$ also include a common random variable vector ${{\boldsymbol{e}}_k}$.
As $K$ increases, the sum of covariances between ${Y_{ik}}$ and ${Y_{jk}}$ $\forall i\ne j$ becomes dominant and the covariances between $X_k$, $\sum\nolimits_j {{\rho _j}{Y_{jk}}} $, and $Z_k$ becomes negligible.  
Then, the variance of $I_k$ can be asymptotically obtained based on Lemmas 1--3 as
$\sigma _{{I_k}}^2/M^2 - \bar \sigma _{{I_k}}^2/M^2\xrightarrow[M,K \to \infty ]{} 0$, where
\begin{equation}
\bar \sigma _{{I_k}}^2 = \rho _k^2\tau _k^4\bar q_k^2 + \tau _k^2\left( {2 - \tau _k^2} \right){\bar q_k} + \sum\nolimits_{j \ne k}^K {\rho _j^2\bar \sigma _{{Y_{jk}}}^2}  + \sum\nolimits_{{i,j\ne k}:{i\ne j}}^K {\rho _i\rho _j{{\bar \omega }_{ijk}}}. 
\end{equation}
${\bar \omega _{ijk}}$ denotes the asymptotic value of ${\omega _{ijk}} = {\rm{Cov}}\left[ {{Y_{ik}},{Y_{jk}}} \right] = {\mu _{{Y_{ik}}{Y_{jk}}}} - {\mu _{{Y_{ik}}}}{\mu _{{Y_{jk}}}}$ $\forall i\ne j$, where ${\mu _{{Y_{ik}}{Y_{jk}}}} = {\rm{E}}\left[ {{Y_{ik}}{Y_{jk}}} \right]$, ${\mu _{{Y_{ik}}}} = {\rm{E}}\left[ {{Y_{ik}}} \right]$, and ${\mu _{{Y_{jk}}}} = {\rm{E}}\left[ {{Y_{jk}}} \right]$.
For convenience, we use the following notations.
\begin{align}
{c_{tk}} & = {\sqrt {1 - \tau _k^2} {\boldsymbol{h}}_{kk}^{\rm{H}}{{\boldsymbol{h}}_{tk}}}
 = \sqrt {\frac{{1 - \tau _k^2}}{{{\kappa _{tk}} + 1}}} \left( {\sqrt {{\kappa _{tk}}} {\boldsymbol{h}}_{kk}^{\rm{H}}{\boldsymbol{h}}_{tk}^{\rm{L}} + \sum\nolimits_p^P {{\boldsymbol{h}}_{kk}^{\rm{H}}{{\boldsymbol{r}}_{tkp}}{g_{tkp}}} } \right),\\
{a_{tmk}} &= \sqrt {\frac{{\tau _k^2}}{{{\kappa _{tk}} + 1}}} \beta _{km}^{\rm{L}}\left( {\sqrt {{\kappa _{tk}}} \beta _{tm}^{\rm{L}}{h_{tkm}} + \sum\nolimits_p^P {{r_{tkmp}}{g_{tkp}}} } \right) ,
\end{align}
for $t \in \left\{ {i,j} \right\}$ and $\forall m,k$.
With ${a_{tmk}}$, we have 
${\tau _k}{\boldsymbol{e}}_k^{\rm{H}}{{\boldsymbol{h}}_{tk}}= \sum\nolimits_m^M {{a_{tmk}}e_{km}^*}$.
Here, ${c_{tk}}$ and ${a_{tmk}}$ are complex Gaussian random variables independent with $e_{km}$ such as ${c_{tk}} \sim \mathcal{CN}\left( {{\mu _{{c_{tk}}}},\sigma _{{c_{tk}}}^2} \right)$ and ${a_{tmk}} \sim \mathcal{CN}\left( {{\mu _{{a_{tmk}}}},\sigma _{{a_{tmk}}}^2} \right)$, where
${\mu _{{c_{tk}}}} = \sqrt {\frac{{{\kappa _{tk}}\left( {1 - \tau _k^2} \right)}}{{{\kappa _{tk}} + 1}}} {\boldsymbol{h}}_{kk}^{\rm{H}}{\boldsymbol{h}}_{tk}^{\rm{L}}$,
$\sigma _{{c_{tk}}}^2 = \frac{{1 - \tau _k^2}}{{{\kappa _{tk}} + 1}}{\sum\nolimits_p^P {\left| {{\boldsymbol{h}}_{kk}^{\rm{H}}{{\boldsymbol{r}}_{tkp}}} \right|} ^2}$,
${\mu _{{a_{tmk}}}} = \sqrt {\frac{{\tau _k^2{\kappa _{tk}}}}{{{\kappa _{tk}} + 1}}} \beta _{km}^{\rm{L}}\beta _{tm}^{\rm{L}}{h_{tkm}}$, and
$\sigma _{{a_{tmk}}}^2 = \frac{{\tau _k^2{{\left( {\beta _{km}^{\rm{L}}} \right)}^2}}}{{{\kappa _{tk}} + 1}}\sum\nolimits_p^P {{{\left| {{r_{tkmp}}} \right|}^2}}.$

By using the above notations, we have 
\begin{align}
{Y_{tk}} &= {\left| {\sqrt {1 - \tau _k^2} {\boldsymbol{h}}_{kk}^{\rm{H}}{{\boldsymbol{h}}_{tk}} + {\tau _k}{\boldsymbol{e}}_k^{\rm{H}}{{\boldsymbol{h}}_{tk}}} \right|^2}
 ={\left| {{c_{tk}}} + \sum\nolimits_m^M {{a_{tmk}}e_{km}^*} \right|^2} \nonumber\\
 &= {\left| {{c_{tk}}} \right|^2} + 2{\mathop{\rm Re}\nolimits} \left( {{c_{tk}}\sum\nolimits_m^M {a_{tmk}^*{e_{km}}} } \right) + \sum\nolimits_m^M {{a_{tmk}}e_{km}^*} \sum\nolimits_m^M {a_{tmk}^*{e_{km}}}. 
\end{align}
With a standard complex Gaussian distribution for ${{e_{km}}}$, we have ${\rm{E}}\left[ {{e_{km}}} \right] = 0$, ${\rm{E}}\left[ {e_{km}^2} \right] = 0$, ${\rm{E}}\left[ {e_{km}^3} \right] = 0$, and ${\rm{E}}\left[ {{{\left| {{e_{km}}} \right|}^2}} \right] = 1$.
Then, ${\mu _{{Y_{tk}}}}$ and ${\mu _{{Y_{ik}}{Y_{jk}}}}$ are obtained as
${\mu _{{Y_{tk}}}} = {\mu _{{C_{tk}}}} + \sum\nolimits_m^M {{\mu _{{A_{tmk}}}}}$ and
\begin{align}
{\mu _{{Y_{ik}}{Y_{jk}}}} &= {\rm{E}}\left[ {{{\left| {{c_{ik}}} \right|}^2}{{\left| {{c_{jk}}} \right|}^2} + {{\left| {{c_{ik}}} \right|}^2}\sum\nolimits_m^M {{{\left| {{a_{jmk}}} \right|}^2}{{\left| {{e_{km}}} \right|}^2}}}\right.\nonumber\\
&\qquad\qquad\qquad + \left.{{\left| {{c_{jk}}} \right|}^2}\sum\nolimits_m^M {{{\left| {{a_{imk}}} \right|}^2}{{\left| {{e_{km}}} \right|}^2}}  + 2{\mathop{\rm Re}\nolimits} \left( {{c_{ik}}c_{jk}^*\sum\nolimits_m^M {a_{imk}^*{a_{jmk}}{{\left| {{e_{km}}} \right|}^2}} } \right) + \xi  \right] \nonumber\\
&= {\mu _{{C_{ik}}}}{\mu _{{C_{jk}}}} {+} {\mu _{{C_{ik}}}}\sum\nolimits_m^M {{\mu _{{A_{jmk}}}}}  {+} {\mu _{{C_{jk}}}}\sum\nolimits_m^M {{\mu _{{A_{imk}}}}}  {+} 2{{\rm Re}} \left( {{\mu _{{c_{ik}}}}{\mu _{c_{jk}}^*}\sum\nolimits_m^M {{\mu _{a_{imk}}^*}{\mu _{{a_{jmk}}}}} } \right) {+}{\mu _\xi },
\end{align}
where we define ${\mu _{{C_{tk}}}} = {\rm{E}}\left[ {{{\left| {{c_{tk}}} \right|}^2}} \right]$ and ${\mu _{{A_{tmk}}}} = {\rm{E}}\left[ {{{\left| {{a_{tmk}}} \right|}^2}} \right]$ for $t \in \left\{ {i,j} \right\}$ and $\forall m,k$.
Then, we have
${\mu _{{C_{tk}}}}= \sigma _{{c_{tk}}}^2 + {\left| {{\mu _{{c_{tk}}}}} \right|^2}$ and
${\mu _{{A_{tmk}}}} = \sigma _{{a_{tmk}}}^2 + {\left| {{\mu _{{a_{tmk}}}}} \right|^2}$.
Further, $\xi $ and ${\mu _\xi }$ are given by
\begin{gather}
\xi  = \sum\limits_m^M {{a_{imk}}e_{km}^*} \sum\limits_m^M {a_{imk}^*{e_{km}}} \sum\limits_m^M {{a_{jmk}}e_{km}^*} \sum\limits_m^M {a_{jmk}^*{e_{km}}} ,\\
{\mu _\xi } = {\rm{E}}\left[\sum\limits_m^M {{{\left| {{a_{imk}}} \right|}^2}{{\left| {{e_{km}}} \right|}^2}} \sum\limits_m^M {{{\left| {{a_{jmk}}} \right|}^2}{{\left| {{e_{km}}} \right|}^2}}  + \sum\limits_m^M {{a_{imk}}a_{jmk}^*{{\left| {{e_{km}}} \right|}^2}} \sum\limits_{n \ne m}^M {a_{ink}^*{a_{jnk}}{{\left| {{e_{kn}}} \right|}^2}}\right]. 
\end{gather}
Then, ${\omega _{ijk}}$ is obtained by
\begin{equation}
{\omega _{ijk}} = 2{\mathop{\rm Re}\nolimits} \left( {{\mu _{{c_{ik}}}}{\mu _{c_{jk}}^*}\sum\limits_m^M {{\mu _{a_{imk}}^*}{\mu _{{a_{jmk}}}}} } \right) + {\mu _\xi }  - \sum\limits_m^M {{\mu _{{A_{imk}}}}} \sum\limits_m^M {{\mu _{{A_{jmk}}}}} .\label{eq.D.Cov2}
\end{equation}
Here, ${\mu _{{c_{tk}}}}$ is obtained from ${\boldsymbol{h}}_{kk}^{\rm{H}}{\boldsymbol{h}}_{tk}^{\rm{L}}$ which is calculated by the sum of $M$ elements.
Then, the first term of ${\omega _{ijk}}$ in (\ref{eq.D.Cov2}) increases with $\mathcal{O}(M^3)$ as $M\to\infty$.
Similarly, the second and the last term of ${\omega _{ijk}}$ increase with $\mathcal{O}(M^2)$ as $M\to\infty$.
According to the scaling laws for $M$, we have
${\omega _{ijk}}/M^2 - {\bar\omega _{ijk}}/M^2 \xrightarrow[M \to \infty ]{} 0$, 
where
\begin{equation}
{\bar\omega _{ijk}} = 2{\mathop{\rm Re}\nolimits} \left( {{\mu _{{c_{ik}}}}{\mu _{c_{jk}}^*}\sum\limits_m^M {{\mu _{a_{imk}}^*}{\mu _{{a_{jmk}}}}} } \right). \label{eq.D.Cov}
\end{equation}
We can observe from (\ref{eq.D.Cov}) that ${\bar\omega _{ijk}}$ can be obtained by a deterministic value and the LOS component of the interference channel exclusively produces ${\bar \omega _{ijk}}$.

\bibliographystyle{IEEEtran}
\bibliography{IEEEabrv,myBiB}

\end{document}